\documentclass[11pt,titlepage,fleqn]{article}
\usepackage{setspace}
\usepackage{framed,subcaption}
\usepackage{url}

\usepackage{amsmath}
\usepackage{latexsym}
\usepackage[dvips]{graphicx,color}
\usepackage{graphicx,color}
\usepackage[usenames,dvipsnames,svgnames,table]{xcolor}
\usepackage{bm}
\usepackage{bbm,amssymb,epsfig}
\usepackage{float}
\usepackage{enumitem}

\usepackage[round]{natbib}

\usepackage{lipsum}

\usepackage{colortbl}
\usepackage{multirow}
\usepackage{graphicx,psfrag,epsfig,epstopdf-base}
\usepackage{natbib}
\usepackage{comment}
\usepackage{wrapfig}
\usepackage{tikz}
\usepackage{geometry}
\usepackage{pdflscape}
\usepackage{epsfig}
\usepackage{graphicx}
\usepackage{amsfonts}
\usepackage{hyperref}
\usepackage{latexsym}
\usepackage{epstopdf}
\usepackage{mathtools}
\usepackage{amsmath,amssymb,amsthm}
\usepackage{bm}
\usepackage{bbm}
\usepackage{booktabs}
\usepackage{parskip}

\newcommand{\bssigma}{\boldsymbol{\Sigma}}
\newcommand{\bTh}{\boldsymbol{\Theta}}
\newcommand{\Del}{\boldsymbol{\Delta}}
\newcommand{\del}{\boldsymbol{\delta}}
\newcommand{\bLam}{\boldsymbol{\Lambda}}

\newcommand{\x}{\mbox{\boldmath $x$}}
\newcommand{\m}{\mbox{\boldmath $m$}}
\newcommand{\bb}{\mbox{\boldmath $b$}}

\newcommand{\W}{\mbox{\boldmath $W$}}
\newcommand{\A}{\mbox{\boldmath $A$}}
\newcommand{\F}{\mbox{\boldmath $F$}}

\newcommand{\U}{\mbox{\boldmath $U$}}

\newcommand{\C}{\mbox{\boldmath $C$}}
\newcommand{\D}{\mbox{\boldmath $D$}}

\newcommand{\R}{\mbox{\boldmath $R$}}
\newcommand{\I}{\mbox{\boldmath $I$}}
\newcommand{\bS}{\mbox{\boldmath $S$}}
\newcommand{\B}{\mbox{\boldmath $B$}}
\newcommand{\f}{\mbox{\boldmath $f$}}
\newcommand{\e}{\mbox{\boldmath $e$}}

\newcommand{\bfepsilon}{\mbox{\boldmath $\epsilon$}}
\newcommand{\ep}{\mbox{\boldmath $\epsilon$}}

\newcommand{\bfzero}{\mbox{\boldmath $0$}}

\newcommand{\code}[1]{\texttt{#1}}

\RequirePackage[OT1]{fontenc}

\graphicspath{{figures/}}
\begin{document}

\title{Efficient Bayesian PARCOR Approaches for Dynamic Modeling of Multivariate Time Series} 

\author{Wenjie Zhao$^{1}$ \and Raquel Prado$^{1}$}
\footnotetext[1]{Department of Statistics, University of California at Santa Cruz. The authors 
were partially supported by NSF award SES-1853210.}

\maketitle
\begin{abstract}
A  Bayesian lattice filtering and smoothing approach is proposed for fast
and accurate modeling and inference in multivariate non-stationary time series.
This approach offers computational feasibility and interpretable time-frequency analysis in
the multivariate context.  The proposed framework allows us to obtain posterior estimates 
of the time-varying spectral densities of individual time series components, 
as well as posterior measurements of the time-frequency relationships
across multiple components, such as time-varying coherence and partial coherence.

The proposed formulation considers multivariate dynamic linear models (MDLMs)
on the forward and backward time-varying partial autocorrelation coefficients
(TV-VPARCOR).  Computationally expensive schemes 
for posterior inference on the multivariate dynamic PARCOR model are avoided using approximations in the
MDLM context.  Approximate inference on the corresponding time-varying vector autoregressive (TV-VAR) coefficients
is obtained via Whittle's algorithm. 
A key aspect of the proposed TV-VPARCOR  representations is that they are 
of lower dimension, and therefore more efficient, than TV-VAR representations.
The performance of the TV-VPARCOR models is illustrated
in  simulation studies and in the analysis of multivariate non-stationary temporal
data arising in neuroscience and environmental applications. 
Model performance is evaluated using goodness-of-fit measurements in the time-frequency domain
and also by assessing the quality of short-term forecasting.


\end{abstract}

\newpage

\allowdisplaybreaks[2]
\setcounter{page}{1}

\section{Introduction} 

Recent technological advances in scientific areas have led to multi-dimensional datasets with a complex temporal structure, 
often consisting of several time series components that are related over time. Inferring changes in the spectral content 
of each component, as well as time-varying relationships across components, is often relevant in applied areas. 
For example, understanding the interplay across temporal components derived from multi-channel/multi-location brain signals
and brain imaging data is a key feature in brain connectivity studies 
\cite[e.g.,][among others]{yuetal2016,chiangetal2017,tingetal,grangerconnectivity}. Multivariate time series
analysis is also important for filtering, smoothing
and prediction in environmental studies and finance where many variables are simultaneously measured over time 
\citep[e.g.,][]{zhangbook,tsay}. 

Several time-domain, frequency-domain and time-frequency approaches are available for modeling and inferring  
spectral characteristics of univariate non-stationary time series. However, a much more
limited number of approaches are available 
for computationally efficient and scientifically interpretable analysis of multivariate non-stationary time series. 
Furthermore, currently available statistical tools have important practical limitations. 
For instance, vector autoregressions (VARs) are often used in the analysis of multi-channel EEG (electroencephalogram)
data, however, these models cannot capture the time-varying characteristics of these data. Alternative approaches
incorporate more flexible and realistic dynamic structures \citep[e.g.,][]{west1999evaluation,prado2001multichannel,nakajimawesteeg},
but are either not available for multivariate time series, or they are highly computationally intensive, requiring Markov
chain Monte Carlo (MCMC) sampling for posterior inference. Frequency-domain and time-frequency approaches
have also been developed, but typically these methods are only able to handle multiple (not multivariate) stationary
time series \citep[e.g.,][]{cadonna2019}, or are methodologically adequate and flexible \citep[e.g.,][]{krafty1,krafty2}, but computationally unfeasible to jointly analyze more than a relatively small number of
multivariate time series components.

In the univariate context \cite{yang2016bayesian} considers a Bayesian lattice filter approach for analyzing a single time series which 
uses univariate dynamic linear models (DLMs) to describe the evolution of the forward and backward
partial autocorrelation coefficients of such series. A key feature of this approach is its computational appeal. A DLM representation of a univariate time-varying autoregression requires a model with a state-parameter vector of dimension $P,$ where $P$ is the order of the autoregression \citep[see e.g.,][]{west1999evaluation,prado2010time}. Therefore, filtering and smoothing in this setting requires the inversion of $P \times P$ matrices at each time $t.$ Alternatively, the DLM formulation in the PARCOR domain requires fitting $ 2 \times P$  DLMs, $P$ for the forward coefficients and $P$ for the backward coefficients, where each DLM  has a  univariate  state-space parameter,  fully avoiding matrix inversions and resulting in computational savings for cases in which $P \geq 2$. 

In this paper we extend the Bayesian lattice filter approach of \cite{yang2016bayesian} to the multivariate case. Our proposed models offer several advantages over currently available multivariate approaches for non-stationary time series including
computational feasibility for joint analysis of relatively large-dimensional multivariate time series, and interpretable 
time-frequency analysis in the multivariate context. In particular, the proposed framework leads to posterior 
estimates of the time-varying spectral densities of each individual time series, as well as posterior measurements of the 
time-frequency relationships across multiple time series over time, such 
as time-varying coherence and partial coherence. 
We note that extending the approach \cite{yang2016bayesian} to the 
multivariate case is non-trivial, as the closed-form inference used in the univariate DLM formulation of 
the lattice filter is not available for the multivariate case considered here. 
Multivariate DLM theory \citep{west1999bayesian,prado2010time} allows for full posterior 
inference in closed-form only when the covariance matrices of the innovations at the observation level and those at the system level are known, which is rarely the case in practice. 
Full posterior inference via MCMC can be obtained for more general multivariate DLM settings, but such posterior sampling schemes 
are very computationally expensive, making them only feasible when dealing with a small number of time series of small/moderate 
time lengths, and low-order TV-VAR models. 
We address these challenges by approximating the covariance matrices of the innovations at the observational level for the multivariate dynamic forward and backward PARCOR models  
using the approach of \cite{triantafyllopoulos2007covariance}. In addition, we use discount factors to specify the structure of the 
covariance matrices at the system levels. 
Our framework casts the time-varying multivariate representation 
of the input-output relations between the vectorial forward and backward predictions of a 
multivariate time series process --and their corresponding forward and backward matrices of PARCOR (partial autocorrelation) coefficients-- as a
Bayesian multivariate state-space model. 
Once approximate posterior inference is obtained for the multivariate time-varying PARCOR 
coefficients, posterior estimates for the implied
time-varying vector autoregressive (TV-VAR) coefficient matrices and innovations covariance matrices can be obtained via Whittle's algorithm \citep{zhou1992algorithms}. Similarly, 
posterior estimates for any 
function of such matrices, such as the multivariate spectra and functions of the spectra, can also be obtained. 
A key feature of the proposed TV-VPARCOR representation is that it is more parsimonious and 
flexible than directly working with the TV-VAR state-space representation. We illustrate this in the 
analyses of simulated and real data 
presented in Sections 3 and 4.  We also propose a method for selecting the number of stages in the 
TV-VPARCOR setting based on an approximate calculation of the Deviance Information Criterion (DIC). 

The paper is organized as follows. Section 2 presents the models and discusses approximate posterior inference. 
Section 3 illustrates the performance of the proposed TV-VPARCOR models in simulation studies. Comparisons with results obtained from DLM representations of TV-VAR models are also provided. 
Section 4 presents analyses of two real multivariate temporal datasets. The first application considers joint analysis of multi-channel electroencephalogram data and the second one illustrates the analysis and forecasting of multi-location bivariate wind components. Finally, Section 5 presents a discussion and briefly describes 
potential future developments.

\section{Models and Methods for Posterior Inference}

\subsection{Time-Varying Vector Autoregressive Models (TV-VAR) and Lattice Filters}
Let $\x_t$ be a $K \times 1$ vector time series for $t= 1, \dots, T.$ A time-varying vector autoregressive model of order $P,$ referred to as TV-VAR$(P),$ is given by 
\begin{align*}
  \x_t = \A_{t,1}^{(P)} \x_{t-1} + \dots + \A_{t,P}^{(P)}\x_{t-P} + \bfepsilon_t, \quad \bfepsilon_t \sim \mathcal{N}(\bfzero, 
	\bssigma_t), 
\end{align*}
where $\A_{t,j}^{(P)}$ is the $K \times K$ matrix of time-varying coefficients at lag $j,$ $j = 1, \dots, P,$ and $\bssigma_t$ 
is the $K\times K$ innovations variance-covariance matrix at time $t.$ The $\bfepsilon_t$s are assumed to be independent over time. 

\cite{yang2016bayesian} considers a Bayesian lattice filter dynamic linear modeling approach 
for the case of univariate time-varying autoregressions (TVAR), 
i.e., when $K=1$ above. Such approach is based on a lattice structure formulation of the univariate Durbin-Levinson 
algorithm \citep[see, e.g.,][]{brockwelldavis,shumway2006time} used in \cite{kitagawa2010}. 
The idea is to obtain posterior estimation on the forward and
backward time-varying PARCOR coefficients using a computationally efficient lattice filter representation. Once dynamic PARCOR estimation is obtained, estimates
of the TVAR coefficients can be derived using the Durbin-Levinson recursion. The main advantage of using the dynamic PARCOR lattice filter representation instead of a dynamic linear model TVAR representation such as that used in \cite{west1999evaluation}, is that the former avoids the inversion of $P \times P$ matrices required in the TVAR DLM filtering and smoothing equations. Instead, the PARCOR approach  
considers $2 P$ dynamic linear models with univariate state parameters (e.g., $P$ DLMs with univariate state parameters for the forward coefficients and $P$ DLMs with univariate state parameters for the backward coefficients), completely avoiding matrix inversions.  This is important for 
computational efficiency when considering models with $P>2$ 
and large $T$.  The PARCOR approach also offers additional modeling advantages due to the fact that considering $2 P$ DLMs with univariate state parameters 
generally provides more flexibility,  than using a single DLM TVAR with $P$-dimensional state parameters. 

We extend the approach of \cite{yang2016bayesian} to consider multivariate non-stationary time series. More specifically, we consider Bayesian multivariate DLMs that use the multivariate Whittle algorithm \citep{zhou1992algorithms}, also known as the multivariate Durbin-Levinson algorithm 
\citep{brockwelldavis}, to obtain a representation of the TV-VAR coefficient matrices in terms of time-varying PARCOR matrices as follows. 
Let $\f_t^{(P)}$ and $\bb_t^{(P)}$ be the $K$-dimensional prediction error vectors at time $t$ for the forward and backward TV-VAR$(P)$ model, respectively, where,
\begin{align*}
  \f^{(P)}_t = \x_t - \sum^P_{j=1}\bm{A}_{t, j}^{(P)}\x_{t-j}, \;\;\; \mbox{and} \;\;\; \bb^{(P)}_t = \x_t - \sum^P_{j=1}\D_{t, j}^{(P)}\x_{t+j}. 
\end{align*}
$\A_{t,j}^{(P)}$ and $\D_{t,j}^{(P)}$ denote, respectively, the $K \times K$ time-varying matrices of forward and backward TV-VAR$(P)$ coefficients for $j=1,\ldots,P.$
Similarly, $\A_{t,j}^{(m)}$ and $\D_{t,j}^{(m)}$ denote the time-varying matrices of forward and backward TV-VAR$(m)$ coefficients for $j=1,\ldots,m.$
Then, we write the $m$-stage of the lattice filter in terms of the pair of input-output relations between the forward and backward $K$-dimensional vector predictions, as follows, 
\begin{align}
  \f_t^{(m-1)} &= \bLam_{t, m}^{(m)}\bb_{t-m}^{(m-1)} + \f_t^{(m)}, \qquad \f_t^{(m)} \sim \mathcal{N}(\bfzero, \bssigma_{t, f, m}), \label{eqn1}\\
  \bb_t^{(m-1)} &= \bTh_{t, m}^{(m)}\f_{t+m}^{(m-1)} + \bb_t^{(m)}, \qquad \bb_t^{(m)} \sim \mathcal{N}(\bm{0}, \bssigma_{t, b, m}), \label{eqn2}
\end{align}
where $\bLam_{t, m}^{(m)}$ and $\bTh_{t, m}^{(m)}$ are, respectively, the $K \times K$ matrices of time-varying forward and backward PARCOR coefficients for $m=1,\ldots,P.$ Note that for stationary AR$(P)$, i.e., models with $K=1$ and static AR coefficients in the stationary region, the forward and backward PARCOR coefficients are equal, i.e., $\lambda_{m}^{(m)}=\theta_{m}^{(m)}$ for all $m.$ For general $K$ and non-stationary processes the forward and backward PARCOR coefficients are not the same. 

For each stage $m$ of the lattice structure above, we obtain the forward and backward TV-VAR coefficient matrices, $\A_{t,m}^{(P)}$ and $\D_{t,m}^{(P)},$ from the time-varying forward and backward PARCOR coefficient matrices, $\bLam_{t, m}^{(m)}$ and $\bTh_{t, m}^{(m)}$, using Whittle's algorithm \citep[see, e.g.,][]{zhou1992algorithms}, i.e., 
\begin{align}
  \A_{t, j}^{(m)} &= \A_{t, j}^{(m-1)} - \A_{t, m}^{(m)}\D_{t, m-j}^{(m-1)}, \label{eqn_lattice_1} \\
  \D_{t, j}^{(m)} &= \D_{t, j}^{(m-1)} - \D_{t, m}^{(m)}\A_{t, m-j}^{(m-1)}, \quad j = 1, \dots, m-1, \label{eqn_lattice_2}
\end{align}
with $\A_{t, m}^{(m)} = \bLam_{t, m}^{(m)}$ and $\D_{t, m}^{(m)} = \bTh_{t, m}^{(m)},$ for $m=1,\ldots,P.$ 

\subsection{Model specification and inference}
Our proposed model specification uses equations (\ref{eqn1}) and (\ref{eqn2}) as observational level equations of multivariate DLMs \citep{west1999bayesian, prado2010time} on the forward and backward PARCOR time-varying coefficients. These multivariate DLMs are specified as follows. For each $t$, let $\mbox{vec}(\bLam^{(m)}_{t, m})$ and $\mbox{vec}(\bTh^{(m)}_{t, m})$ be the vectorized forward and backward PARCOR coefficients, i.e., these are $K^2$ vectors obtained by stacking the forward and backward PARCOR coefficient matrices at time $t$, $\bLam^{(m)}_{t, m}$ and $\bTh^{(m)}_{t, m},$ by columns, respectively. In addition, define the forward and backward $K \times K^2$ matrices $\F^{(m-1)}_{t+m} =  (\f_{t+m}^{(m-1)}) \otimes \I_{K \times K} $ and $\B^{(m-1)}_{t-m}= (\bb_{t-m}^{(m-1)}) \otimes \I_{K \times K},$ where $\I_{K \times K}$ denotes the $K \times K$ identity matrix and $\otimes$ denotes the Kronecker product. Then, equations (\ref{eqn1}) and (\ref{eqn2}) can be rewritten as
\begin{align}
  \f_t^{(m-1)} &= \B^{(m-1)}_{t-m}\mbox{vec}(\bLam_{t, m}^{(m)}) + \f_t^{(m)}, \quad \f_t^{(m)} \sim \mathcal{N}(\bm{0}, \bssigma_{t, f, m}), \label{eqn1new} \\
  \bb_t^{(m-1)} &= \F^{(m-1)}_{t+m}\mbox{vec}(\bTh_{t, m}^{(m)}) + \bb_t^{(m)}, \quad \bb_t^{(m)} \sim \mathcal{N}(\bm{0}, \bssigma_{t, b, m}), \label{eqn2new}
\end{align}
which correspond to the observational equations of two multivariate dynamic linear regressions on $\f_t^{(m-1)}$ and $\bb_t^{(m-1)}$, with 
dynamic coefficients $\mbox{vec}(\bLam_{t,m}^{(m)})$ and $\mbox{vec}(\bTh_{t,m}^{(m)}),$ respectively. In order to complete the MDLM structure we specify random walk evolution equations for $\mbox{vec}(\bLam_{t,m}^{(m)})$ and $\mbox{vec}(\bTh_{t,m}^{(m)})$ as follows, 
\begin{align}
  \mbox{vec}(\bLam_{t, m}^{(m)}) &= \mbox{vec}(\bLam_{t-1, m}^{(m)}) + \ep_{t, f, m}, \quad \ep_{t, f, m} \sim \mathcal{N}(\bm{0}, \W_{t, f, m}), \label{evol_1} \\
   \mbox{vec}(\bTh_{t, m}^{(m)}) &= \mbox{vec}(\bTh_{t-1, m}^{(m)}) + \ep_{t, b, m}, \quad \ep_{t, b, m} \sim \mathcal{N}(\bm{0}, \W_{t, b, m}), \label{evol_2}
\end{align}
where $\W_{t, f, m}$ and $\bm{W}_{t,b, m}$ are time dependent system covariance matrices. Finally, we specify prior distributions for $\mbox{vec}(\bLam_{0,m}^{(m)})$ and $\mbox{vec}({\bTh}_{0,m}^{(m)})$ and all $m.$ We use conjugate normal priors for these parameters, i.e., we assume
\begin{eqnarray}
\mbox{vec}(\bLam_{0,m}^{(m)}) | \mathcal{D}_{0,f,m} & \sim & \mathcal{N}(\m_{0, f, m}, \C_{0, f, m}), \label{prior_f} \\
\mbox{vec}(\bTh_{0,m}^{(m)}) | \mathcal{D}_{0,b,m} & \sim & \mathcal{N}(\m_{0,b, m}, \C_{0,b, m}), \label{prior_b} 
\end{eqnarray}
where $\mathcal{D}_{0,f,m}$ and $\mathcal{D}_{0,b,m}$ denote the information available at time $t=0$ for the forward and backward state parameter vectors, respectively. 

Given $\bssigma_{t,f,m},$ and $\W_{t,f,m}$ for all $t=1,\ldots,T,$ and all $m=1,\ldots,P,$ equations (\ref{eqn1new}), (\ref{evol_1}) and (\ref{prior_f})
define a normal MDLM \citep[see, e.g.,][Chapter 10]{prado2010time} for the forward time-varying PARCOR. Similarly, given 
$\bssigma_{t,b,m}$ and $\W_{t,b,m}$ for all $t$ and all $m,$ equations
(\ref{eqn2new}), (\ref{evol_2}) and (\ref{prior_b}) define a normal MDLM for the backward time-varying PARCOR.

Note that posterior inference in the case of univariate models with $K=1$ is available in closed form via the DLM filtering and smoothing equations. 
This is used in \cite{yang2016bayesian} to obtain posterior inference in this univariate case. 
However, posterior inference in the general multivariate setting proposed here is not 
available in closed form when the observational and system covariance matrices are unknown, which is typically the case in 
practical settings. Therefore, as explained below, we use discount factors to specify $\W_{t,f,m},$ and $\W_{t,b,m}.$ We also assume $\bssigma_{t,f,m}=\bssigma_{f,m}$ and $\bssigma_{t,b,m}=\bssigma_{b,m}$ for
all $t,$ and use the approach of \cite{triantafyllopoulos2007covariance} to obtain estimates of $\bssigma_{f,m}$ and $\bssigma_{b,m},$ which allows us 
to get approximate posterior inference in the multivariate case. 

We first define the $K^2 \times K^2$ system covariance matrices using discount factors by setting
$$\Del_{f, m} = diag(\del_{f, m, 1}^{-1/2}, \dots, \del_{f, m, K}^{-1/2}), \;\;\; \mbox{and} \;\;\; \Del_{b, m} = diag(\del_{b, m, 1}^{-1/2}, \dots, \del_{b, m, K}^{-1/2}),$$ where each component,  $\del_{., m, i},$ is a $K$-dimensional  vector that contains the discount factors for each of the $K$ components at stage $m$. Although we can assume different discount factors for different elements of $\del_{\cdot,m,k}$ and also across different $k$s, in practice we usually set all the elements of $\del_{f,m,k}$ equal to
$\delta_{f,m}$ and all the elements of $\del_{b,m,k}$ equal to $\delta_{b,m}$ for all $k=1,\ldots,K,$  and then choose $\delta_{f,m}$ and $\delta_{b,m}$ optimally according to some criterion for each stage $m$ (this is discussed in Section 2.3). This structure for $\Del_{f,m}$ and $\Del_{b,m}$ allows us to obtain closed form expressions for $\W_{t,f,m}$ and $\W_{t,b,m}$ sequentially over time. 

We now describe the full algorithm for approximate posterior inference in the forward TV-VPARCOR model. The 
algorithm for the backward model is similar. Let $\mathcal{D}_{t,f,m}$ denote all the information available up to time $t$ at 
stage $m$ for the forward model, with $\mathcal{D}_{t,f,m}=\{ \mathcal{D}_{t-1,f,m}, \f_t^{(m-1)} \}$. 
Consider the posterior 
expectation of $\bssigma_{f,m}$ up to time $t,$ i.e., $E(\bssigma_{f,m}| \mathcal{D}_{t,f,m}),$
and assume that $\lim_{t \rightarrow \infty} E(\bssigma_{f,m}| \mathcal{D}_{t,f,m})=
\bssigma_{f,m}.$ Let $n_{0,f,m}$ be a positive scalar and $\bS_{0,f,m}$ be the prior expectation 
of $\bssigma_{f,m}.$ Assume that at time $t-1,$ we have that $\mbox{vec}(\bLam_{t-1,m}^{(m)}) | \mathcal{D}_{t-1,f,m}$  is approximately distributed as $N(\m_{t-1,f,m}, \C_{t-1,f,m}),$ 
and so, $E(\f_{t}^{(m-1)} | \mathcal{D}_{t-1,f,m})$ is approximated by $\B_{t-m}^{(m-1)} \m_{t-1,f,m}$ and $V(\f_t^{(m-1)} |\mathcal{D}_{t-1,f,m})$ is approximated by  $\bm{Q}_{t-1,f,m}=\B_{t-m}^{(m-1)} \R_{t,f,m} (\B_{t-1}^{(m-1)})' + \bS_{t-1,f,m},$ with $\R_{t,f,m}= \C_{t-1,f,m} + \W_{t,f,m},$ for some $\bS_{t-1,f,m}.$ 
Then, following Theorem 1 of \cite{triantafyllopoulos2007covariance} we have that, if $\bssigma_{f,m}$ is bounded, $\bS_{t,f,m}$
will approximate $\bssigma_{f,m}$ for $t$ large, with 
\begin{eqnarray}
	\bS_{t, f, m} = \frac{1}{(n_{0, f, m} + t)} \left( n_{0, f, m} \bS_{0, f, m} + \sum^t_{i=1} \bS_{i-1,f, m}^{1/2}\bm{Q}_{i, f, m}^{-1/2}\e_{i,f, m}\e_{i,f,m}'\bm{Q}_{i,f, m}^{-1/2}\bS_{i-1,f, m}^{1/2}\right), \label{eqn3}
\end{eqnarray}
where in our case $\e_{t,f,m}=\f_{t}^{(m-1)}-\B_{t-m}^{(m-1)} \m_{t-1,f,m},$ and  $\bS_{i-1,f, m}^{1/2}, \bm{Q}_{i, f, m}^{-1/2}$ are symmetric square roots of the matrices $\bS_{i-1, f, m}$ and $\bm{Q}_{i, f, m}^{-1},$ respectively, based on the spectral decomposition factorization of symmetric positive definite matrices for all $i=1,\ldots,t.$

Using the approximation above we obtain the filtering equations below for approximate inference in the forward TV-VPARCOR model. 
\begin{itemize}
\item[-] The one-step ahead forecast mean and covariance at time $t$ are given by: 
	$$E(\f^{(m-1)}_{t}|\mathcal{D}_{t-1,f,m}) \approx \B_{t-m}^{(m-1)} \m_{t-1,f,m}.$$ 
and 
		$$V(\f^{(m-1)}_{t}|\mathcal{D}_{t-1,f,m}) \approx \bm{Q}_{t,f,m}= \B_{t-m}^{(m-1)}\R_{t, f, m}(\B_{t-m}^{(m-1)})' + 
\bS_{t-1, f, m},$$ 
		where $\R_{t, f, m} = \C_{t-1, f, m} + \W_{t, f, m}$
		and $\W_{t, f, m} = \Delta_{f, m}\C_{t-1, f, m}\Delta_{f, m} - \C_{t-1, f, m}$. 
	\item[-] The one-step forecast error vector is given by $\e_{t, f, m} = \f^{(m-1)}_{t} - \B_{t-m}^{(m-1)} \m_{t-1,f,m}.$ 
\item[-] Using Bayes' theorem and the equations above we can obtain the approximate posterior distribution at time $t$ as
	$\mbox{vec}(\bLam^{(m)}_{t, m})|\mathcal{D}_{t,f,m}  \approx \mathcal{N}(\m_{t, f, m}, \C_{t, f, m}),$
where 
\begin{align}
\m_{t, f, m} &= \m_{t-1, f, m} + \U_{t, f, m} \e_{t, f, m}, \label{filter1}\\
	\C_{t, f, m} &= \Del_{f, m}\C_{t-1, f, m}\Del_{f, m} + \bm{U}_{t, f, m}\bm{Q}_{t, f, m}\bm{U}'_{t, f, m}, \label{filter2}\\ 
\bm{U}_{t, f, m} &= \bm{\Delta}_{f, m}\bm{C}_{t-1, f, m}\bm{\Delta}_{f, m}\bm{B}^{(m-1)}_{t-m}\bm{Q}^{-1}_{t, f, m}. \label{filter3} 
\end{align}
\end{itemize}
Approximate filtering and predictive distributions for 
$\mbox{vec}(\bm{\Lambda}_{t, m}^{m})| \mathcal{D}_{t, f, m}$, $\bm{f}^{(m-1)}_t | \mathcal{D}_{t, f,m}$ and 
$\bm{f}^{(m-1)}_{t+h} | \mathcal{D}_{t, f, m}$ for a positive integer $h > 0$ can also be obtained by taking $\bssigma_{f,m}= \bm{S}_{t,f,m}.$ 

After applying the filtering equations up to time $T$, it is possible to compute approximate smoothing distributions for the forward PARCOR model 
by setting $\bssigma_{f,m}=\bm{S}_{T,f,m}.$ This leads to approximate smoothing distributions 
$$\mbox{vec}(\bm{\Lambda}_{t,m}^{(m)}) | \mathcal{D}_T \approx \mathcal{N}(\bm{a}_{T,f,m}(t-T), \bm{R}_{T,f,m}(t-T)),$$ where 
the mean and covariance are computed recursively via
\begin{eqnarray}
	\bm{a}_{T,f,m}(t-T) & = & \bm{m}_{t,f,m} - \bm{J}_{t,f,m} (\bm{a}_{t+1,f,m} - \bm{a}_{T,f,m}(t-T+1)), \label{smoothing1} \\
        \bm{R}_{T,f,m}(t-T) & = & \bm{C}_{t,f,m} - \bm{J}_{t,f,m} (\bm{R}_{t+1,f,m} - \bm{R}_{T,f,m}(t-T+1)), 
\label{smoothing2} 
\end{eqnarray}
for $t=(T-1),\ldots,1,$ with $\bm{J}_{t,f,m}=\bm{C}_{t,f,m} \bm{R}_{t+1,f,m}^{-1},$ and starting values $\bm{a}_T(0)=\bm{m}_{T,f,m}$ and $\R_{T,f,m}(0)=\bm{C}_{T,f,m}.$ Filtering and smoothing equations can be obtained for the backward PARCOR model in a similar manner. Finally, the algorithm for approximate posterior estimation is as follows. 

\textbf{Algorithm} 
\begin{itemize}
  \item[1.] Given  hyperparameters $\{P, \bm{\Delta}_{f,m}, \bm{\Delta}_{b, m}; m = 1, \dots, P\},$ set $\bm{f}_t^{(0)} = \bm{b}_t^{(0)} = \bm{x}_t,\; \textrm{for } t = 1, \dots, T.$
  \item[2.] Use $\{\bm{f}^{(0)}_t\}$ and $\{\bm{b}^{(0)}_t\}$ as vectors of responses in the observational
level equations (\ref{eqn1}) and (\ref{eqn2}), respectively, which, combined with the random walk evolution
equations (\ref{evol_1}) and (\ref{evol_2}), and the priors (\ref{prior_f}) and (\ref{prior_b}), 
define the multivariate PARCOR forward and backward models. Then, use 
the sequential filtering equations (\ref{filter1}) to (\ref{filter3})  to obtain the estimated $\{\bm{S}_{T, f, 1}\}$ and $\{\bm{S}_{T, b, 1}\}$.   Use the sequential filtering equations (\ref{filter1}) to (\ref{filter3}) along with the smoothing equations (\ref{smoothing1}) and (\ref{smoothing2}) to obtain a series of estimated parameters 
$\{\mbox{vec}(\hat{\bm{\Lambda}}^{(1)}_{t, 1})\}$, $\{\mbox{vec}(\hat{\bm{\Theta}}^{(1)}_{t, 1})\}$ for $t=1:T$. These estimated parameters are set at the posterior means of the smoothing distributions, i.e., the values in (16) for the forward case and a similar equation in the backward case.
\color{black} 
  \item[3.] Use the observational equations (\ref{eqn1}) and (\ref{eqn2}) to obtain the new series of forward and backward prediction errors, $\{\bm{f}_t^{(1)}\}$ and $\{\bm{b}_t^{(1)}\},$ for $t= 1, \dots, T.$ 
  \item[4.] Repeat steps 2-3 above until $\{\mbox{vec}(\hat{\bm{\Lambda}}^{(m)}_{t, m})\}$, $\{\mbox{vec}(\hat{\bm{\Theta}}^{(m)}_{t, m})\}$, 
$\{\bm{S}_{T, f, m}\}$ and $\{\bm{S}_{T, b, m}\}$ have been obtained for all $m=1,\ldots,P.$ 
  \item[5.] Finally, use $\{\mbox{vec}(\hat{\bm{\Lambda}}^{(m)}_{t, m})\}$ and
$\{\mbox{vec}(\hat{\bm{\Theta}}^{(m)}_{t, m})\}$, for $m = 1,\dots,P,$ 
and equations (\ref{eqn_lattice_1}) and 
(\ref{eqn_lattice_2}) to obtain the forward and backward TV-VAR coefficient matrices via Whittle's algorithm. 
\end{itemize}
\color{black}

\subsection{Model selection and time-frequency representation}
In order to select the optimal model order and discount factors, 
we begin by specifying a potential maximum value of $P,$ say $P_{\max},$ for the model order. At level $m$ we search for the optimal values of $\bm{\Delta}_{f, m}$ and $\bm{\Delta}_{b, m}.$ In other words, at level $m=1$ we search for 
the combination of values of $\bm{\Delta}_{f,1}$ and $\bm{\Delta}_{b, 1}$ maximizing the log-likelihood resulting
from $(\ref{eqn1})$ with $m=1$. Using the selected optimal $\bm{\Delta}_{f,1}$ and $\bm{\Delta}_{b,1}$, 
we can obtain the corresponding series $\{\bm{f}_t^{(2)}\}$ and $\{\bm{b}_t^{(2)}\}$, for $t = 1, \dots, T$, as well as the maximum log-likelihood value $\mathcal{L}_{f,1}.$ Then, we repeat the above 
search procedure for stage two, i.e., $m=2$, using the output $\{\bm{f}_t^{(2)}\}$ and $\{\bm{b}_t^{(2)}\}$ obtained from implementing the filtering and smoothing equations with the previously selected hyperparameters $\bm{\Delta}_{f, 1}$ and $\bm{\Delta}_{b, 1}$. We obtain optimal $\bm{\Delta}_{f,2}$, $\bm{\Delta}_{b,2}$ as well as  $\{\bm{f}_t^{(3)}\}$ and $\{\bm{b}_t^{(3)}\}$, 
for $t = 1, \dots, T.$ We also obtain the value of the corresponding maximum log-likelihood $\mathcal{L}_{f,2}$. 
We repeat the procedure until the set $\{\bm{\Delta}_{f, m}, \bm{\Delta}_{b, m}, \mathcal{L}_{f,m}\}, m = 1, \dots P_{\max},$ has been selected. We then consider two different methods for selecting the optimal model order as described below.  
Note that one can also obtain the optimal likelihood values from the backward model, $\mathcal{L}_{b,m},$ for $m=1,\ldots P_{\max}.$ For all the examples and real data analyses presented below we choose the optimal model orders based on the 
optimal likelihood values for the forward model. Similar results were obtained based on the optimal likelihood values for the backward models. 

\noindent{\bf Method 1: Scree plots.} This method was used by \cite{yang2016bayesian} to select the model order visually
by plotting $\mathcal{L}_{f,m}$ against the order $m.$ The idea is that, when the observed vector of time series truly follows a TV-VAR
model, the values of $\mathcal{L}_{f,m}$ will stop increasing after a specific lag and this lag is then chosen to be the model
order. A numerical version of this method can also be implemented by computing the percent of change in the likelihood going from $\mathcal{L}_{f,m-1}$ to $\mathcal{L}_{f,m},$ however, here we use scree plots as a visualization tool and use the model selection criterion below to numerically find an optimal model order. 


\noindent{\bf Method 2: DIC model selection criterion.} We consider an approach based on the deviance information criterion (DIC) to choose the model order
\citep[see][and references therein]{gelman2014bayesian}. 
In general, for a model with parameters denoted as $\bm{\theta},$ the DIC is defined as 
\begin{align*}
    {DIC} &= -2\log p(\bm{y}|\hat{\bm{\theta}}_{Bayes}) + 2p_{DIC},
\end{align*}
where $\bm{y}$ denotes the data, $\hat{\bm{\theta}}_{Bayes}$ is the Bayes estimator of $\bm{\theta}$ and $p_{DIC}$ is the effective number of parameters. The effective number of parameters is
given by  
\begin{align*}
    p_{DIC} &= 2\left[ \log p(\bm{y}|\hat{\bm{\theta}}_{Bayes}) - E_{post}\left( \log p(\bm{y}|\bm{\theta}) \right) \right], 
\end{align*}
where the expectation in the second term is an average of $\bm{\theta}$ over its posterior distribution. The expression above is typically estimated using samples $\bm{\theta}^s, s = 1, \dots, S,$ from the posterior distribution as 
\begin{align*}
    \hat{p}_{DIC} &= 2\left[ \log p(\bm{y}|\hat{\bm{\theta}}_{Bayes}) - \frac{1}{S}\sum^S_{s=1}\log p(\bm{y}|\bm{\theta}^s) \right].
\end{align*}
Note, however, that in our case we do not have samples from the exact posterior distribution of the parameters since we are using approximate  inference to avoid computationally costly exact inference via MCMC. Therefore, for a given model order $m$ we compute the likelihood term in the DIC calculation approximately using the forward filtering distributions as explained below.  Also, note that, fitting a PARCOR model at stage $m$ requires fitting all the models of the previous $m-1$ stages. Therefore, the effective number of parameters at stage $m$ is computed by adding the estimated
effective number of parameters of stage $m$ plus the estimated effective number of parameters
for the previous $m-1$ stages.  
In other words, for each stage $m:$ 
\begin{itemize}
    \item[-] Compute the estimated implied log-likelihood from equation (\ref{eqn1new}) for $t=1,\ldots,T,$ using   ${\mbox{vec}(\hat{\bm{\Lambda}}_{t, m}^{(m)}})$ and $\bm{S}_{T,m,f}$. In this way we obtain the first term in the calculation of the DIC for model order $m.$ 
    \item[-] Obtain samples,  $\mbox{vec}(\bLam^{(m)}_{t, m, s})$, for $s=1,\ldots,S,$ from the approximate
	    sequential filtering equations with distributions $\mathcal{N}(\bm{m}_{t, f, m}, \bm{C}_{t, f, m}),$ and use these samples to compute the estimated number of parameters related only to stage $m$ which we denote as $\hat{p}_{DIC,m}^m$. Note that, as mentioned above, stage $m$ requires fitting all the PARCOR models for the previous $(m-1)$ stages and so, in the final DIC calculation at stage $m$ the total estimated effective number of parameters is computed as 
    $$ \hat{p}_{DIC}^m= \sum_{l=1}^m \hat{p}_{DIC,l}^l.$$
		We denote the final estimated DIC for model order $m$ as $\widehat{DIC}_m.$
\end{itemize}

\subsection{Posterior summaries} 
Once an optimal TV-VPARCOR model is chosen
we can obtain posterior summaries 
of any quantities associated to such
model. For instance, we can obtain posterior summaries of the TV-VPARCOR coefficients over time at each stage,
and consequently summaries of the corresponding TV-VAR$(P)$ coefficients over time.  

Time-frequency representations are generally more useful in practice, and these can be obtained 
by computing the spectral density matrix, $\bm{g}(t,\omega),$ for any time $t$ and frequency $\omega \in (0,1/2),$ as well as measurements derived 
from this matrix such as coherence or partial coherence. 
The spectral density matrix is estimated as 
\begin{align}
	{ \hat{\bm{g}}(t, \omega) }= \bm{\hat{\Phi}}^{-1}(t,\omega) \times \bm{\hat{\Sigma}}  \times \bm{\hat{\Phi}}^*(t,\omega)^{-1}, \label{eqn_sp}
\end{align}
where $\bm{\hat{\Phi}}(t,\omega) = \bm{I} - \sum_{m = 1}^P\bm{\hat{A}}^{(P)}_{t,m}\exp\{-2\pi i m\omega\},$ with $ i = \sqrt{-1}$ \citep[see e.g.,][Chapter~4]{shumway2006time}. 
$\bm{\hat{\Sigma}}$ can be set at $\mathbf{S}_{T,f,P}.$ Note that the spectral density matrix ${\bm{g}}(t, \omega)$ consists of individual spectra ${g}_{j,j}(t, \omega)$  for each  component
$j = 1,\dots,K$ of $\x_t,$ and the cross-spectra ${g}_{i,j}(t,\omega)$ between components $i$ and $j.$ From these we can compute
the estimated squared coherence between components $i$ and $j$ as 
$$\hat{\rho}^2_{i,j}(t, \omega) = |{ \hat{g}_{i,j}}(t, \omega)|^2/\{{\hat{g}}_{i,i}(t, \omega){\hat{g}}_{j,j}(t, \omega)\},$$
for all $i \neq j.$ This measure is used to estimate the power transfer between two components of the time series. 
Similarly, the partial squared coherence between components $i$ and $j$ can be estimated as follows. Let $\bm{c}(t,\omega)={\bm{g}}^{-1}(t,\omega)$  be
the inverse of the spectral density matrix with elements $c_{i,j}(t,\omega)$  for $i,j=1,\dots,K.$ Then, the estimated squared partial coherence between
components $i$ and $j$ is given by 
$$ \hat{\gamma}^2_{i,j}(t,\omega)= | \hat{c}_{i,j}(t,\omega)|^2/\{ \hat{c}_{i,i}(t,\omega) \hat{c}_{j,j}(t,\omega) \}.$$
The squared partial coherence is essentially the frequency domain squared correlation coefficient between components $i$ and $j$ after the 
removal of the linear effects of all the remaining components of $\x_t.$ 
Finally, uncertainty measures for the spectral density matrix, and any functions of this matrix, can be obtained from 
the approximate filtering and smoothing posterior distributions of the forward and backward TV-VPARCOR models.

\subsection{Forecasting}
In this section, we show how to obtain $h$-steps ahead forecasts. In order to have a non-explosive behavior in the forecasts, we assume the series is locally stationary in the future, i.e, $\bLam^{(m)}_{t, m} = \bTh^{(m)}_{t, m}$ at time $t = T+1, \dots, T+h$. Then, the approximate $h$-steps ahead forecast posterior distribution of the PARCOR coefficients,  with $h>0,$ is approximated as $(\bLam_{T+h, m}^{(m)} |\mathcal{D}_{T, f, m}) \approx \mathcal{N}(\bm{m}_{T, f, m}(h), \bm{C}_{T, f, m}(h))$, where 
$$ \bm{m}_{T,f,m}(h) = \bm{m}_{T, f, m}; \qquad \bm{C}_{T,f,m}(h) = \bm{C}_{T, f, m} + h\cdot \bm{W}_{T+1, f, m},$$
with $\bm{W}_{T+1,f,m}= \Delta_{f,m} \C_{T,f,m} \Delta_{f,m} - \C_{T,f,m},$
for $m=1,\ldots,P.$
Then, we  apply Whittle's algorithm to transform the PARCOR coefficients, $\bLam_{T+h, P}^{(P)}$, into TV-VAR coefficients $\bm{A}_{T+h, j}^{(P)}$ and $\bm{D}_{T+h, j}^{(P)}$, for $j = 1, \dots, P.$ Finally, we obtain the $h$-steps ahead forecasts using
\begin{eqnarray*}
\hat{\bm{x}}_{T+h} = \sum_{i=1}^{P}\hat{\bm{A}}_{T+h, i}^{(P)} \hat{\bm{x}}_{T+h-i} + \hat{\bm{\epsilon}}_{T+h}^{(P)},
 \;\;\;\;\;  \hat{\bm{\epsilon}}_{T+h}^{(P)} \sim \mathcal{N}(\bm{0}, \bm{S}_{T,f,P}).
 \end{eqnarray*}

\color{black}

\section{Simulation Studies} 
In this section we illustrate our proposed approach in the analysis of simulated data. The relative performances of 
the models considered here, including that of the proposed TV-VPARCOR, were 
assessed by computing the average squared error (ASE) between the estimated spectral density matrix and the true
spectral density matrix.

\subsection{Bivariate TV-VAR$(2)$ processes}
\begin{figure}[ht] 
  \begin{subfigure}[b]{0.5\linewidth}
    \centering
    \includegraphics[width=0.75\linewidth]{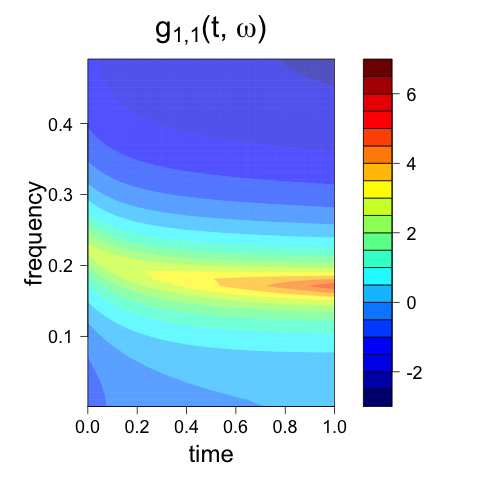} 
    \vspace{4ex}
  \end{subfigure}
  \begin{subfigure}[b]{0.5\linewidth}
    \centering
    \includegraphics[width=0.75\linewidth]{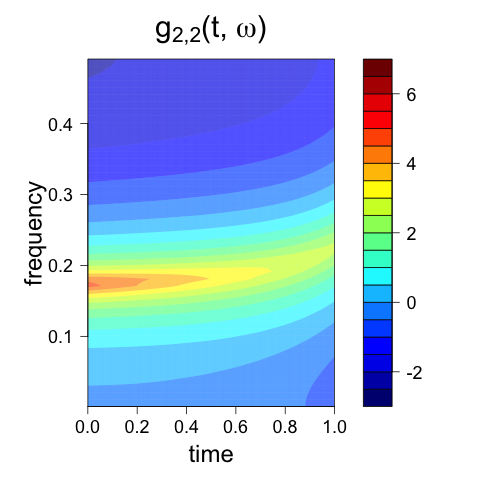} 
    \vspace{4ex}
  \end{subfigure} 
\caption{Case $\phi_{1,1,2} = 0.$ Left: True log spectral density ${g}_{11}(t, \omega)$. Right: True log spectral density ${g}_{22}(t, \omega)$.} \label{fig_ex1}
\end{figure}

We simulated $50$ bivariate time series of length $T=1024$ from the following TV-VAR$(2)$ model:
$$ \bm{x}_t = \bm{\Phi}_{1, t}\bm{x}_{t-1} + \bm{\Phi}_{2, t}\bm{x}_{t-2} + \bm{\epsilon}_t, \quad \bm{\epsilon}_t \sim \mathcal{N}(\bm{0}, \bm{I}_2),$$
with 
\[ \bm{\Phi}_{1, t}=
\begin{pmatrix}
  r_{1, t}\cos(\frac{2\pi}{\lambda_{1, t}}) & \phi_{1,1,2}\\
  0 & r_{2, t}\cos(\frac{2\pi}{\lambda_{2, t}}) \\
\end{pmatrix} \;\;\; \mbox{and} \;\;\; 
\bm{\Phi}_{2, t} = \begin{pmatrix}
-r^2_{1, t} & 0 \\
 0 & -r^2_{2, t}\\
\end{pmatrix}
,\] where 
$r_{1, t} = \frac{0.1}{1024}t + 0.85,$ $r_{2, t} = -\frac{0.1}{1024}t + 0.95,$ 
$\lambda_{1, t} = \frac{15}{1024}t + 5,$ and $\lambda_{2, t} = -\frac{10}{1024}t + 15.$
We also considered two values for $\phi_{1,1,2},$ namely (i) $\phi_{1,1,2} = 0$; 
(ii) $\phi_{1,1,2} = -0.8.$ The true $2 \times 2$ spectral matrix of this process is given by 
\begin{align*}
	{\bm{g}}(t, \omega) = \bm{\Phi}^{-1}(t,\omega) \times \bm{\Sigma} \times {\bm{\Phi}}^*(t,\omega)^{-1}, 
\end{align*}
where $\bm{\Phi}(t,\omega) = \bm{I}_2 - \bm{\Phi}_{1, t}\exp\{-2\pi i\omega\} - \bm{\Phi}_{2, t}\exp\{-4\pi i \omega\}$, and $\bm{\Sigma} = \bm{I}_2.$ The spectral
matrix ${\bm{g}}(t, \omega)$ is  symmetric, with corresponding components ${g}_{11}(t, \omega),$ ${g}_{12}(t, \omega)$ and ${g}_{22}(t, \omega)$, representing, respectively, the spectrum of the first component, the co-spectrum between the first and the second components, and the spectrum of the second component.  The squared coherence between the first and second components is given by 
$$\rho^2_{12}(t,\omega) = \frac{|{g}_{12}(t, \omega)|^2}{{g}_{11}(t, \omega){g}_{22}(t, \omega)}.$$ 
Note that when $\phi_{1, 1, 2} = 0$, the two processes are uncorrelated and ${g}_{1,2}(t,\omega)= \rho_{12}^2(t,\omega)=0$ for all $t$ and $\omega.$  Figure \ref{fig_ex1} shows the true log spectral densities ${g}_{11}(t,\omega)$ and ${g}_{22}(t,\omega).$ 
The true log spectral densities and square coherences for the case of $\phi_{1,1,2}=-0.8$ are
shown in the top row plots of Figure \ref{fig_ex3}. 

We fit bivariate TV-VPARCOR models to each of the 50 simulated bivariate time series under cases (i) and (ii). 
We set a maximum of order $P_{\max} = 5$. The elements of the diagonal component of discount factor matrices $\bm{\Delta}_{f, m}$ 
and $\bm{\Delta}_{b, m},$ $\delta_{f,m}$ and $\delta_{b,m}$ respectively, were chosen from a grid of values in $(0.995, 1)$.  We set the hyperparameters $n_{f, m, 0} = n_{b, m, 0} = 1$, $\bm{S}_{0, f, m} = \bm{S}_{0, b, m} = \bm{I}_2$, $\bm{m}_{0, f, m} = \bm{m}_{0, b, m} = (0, 0, 0, 0)'$ and $\bm{C}_{0, f, m} = \bm{C}_{0, b, m} = \bm{I}_4$. For comparison, we also fit TV-VAR models to the simulated bivariate data with model orders ranging from $1$ to $5$. Multivariate DLM representations of bivariate TV-VAR$(m)$ process were considered for each $m=1,\ldots,5$. Each 
TV-VAR representation has an $(m\times 4)$-dimensional state parameter vector. For each model order a single optimal discount factor, $\delta_m$ was chosen from a grid of values in $(0.995, 1).$ Furthermore, in order to provide a similar model setting to the one we used in our TV-VPARCOR approach, the covariance matrix at the observational level in the DLM formulation for each TV-VAR$(m)$ was also specified following the approach of \cite{triantafyllopoulos2007covariance}. 

\begin{figure}[t]
\centering
\includegraphics[width=5.3cm]{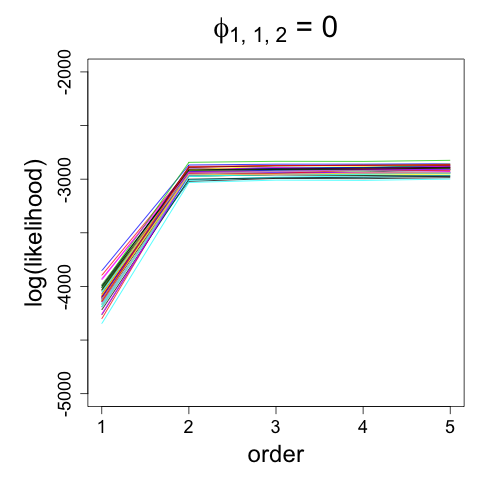}
\includegraphics[width=5.3cm]{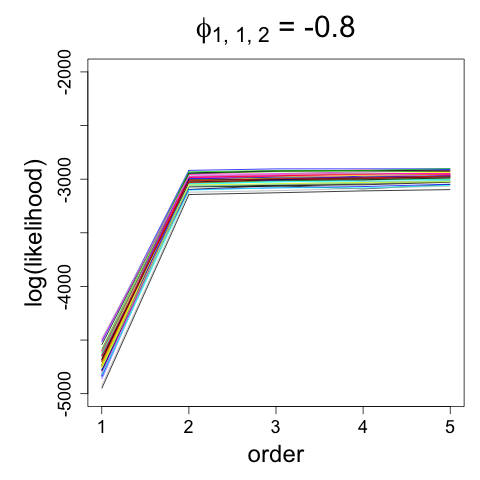}\hfill\\
\includegraphics[width=5.3cm]{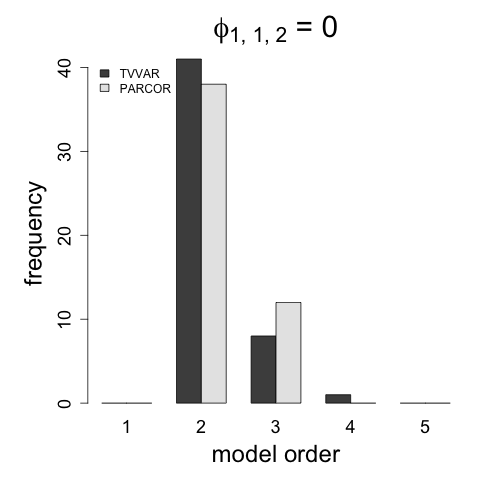}
\includegraphics[width=5.3cm]{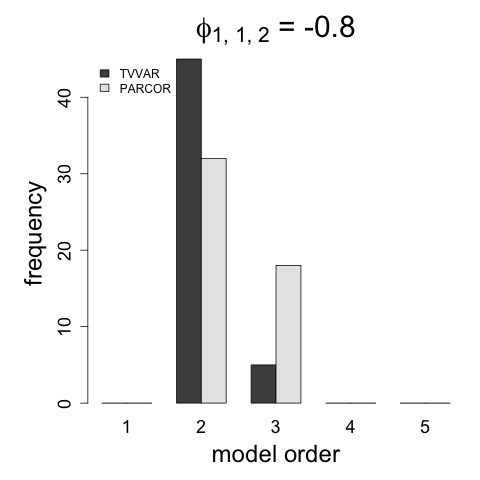}\hfill
	\caption{Top: BLF-scree plots of the $50$ realizations of the TV-VAR(2) process for $\phi_{1, 1, 2} = 0$ and 
	 $\phi_{1, 1, 2} = -0.8$ . Bottom: Optimal model orders for $\phi_{1, 1, 2} = 0$ and 
	 $\phi_{1, 1, 2} = -0.8$.}  \label{fig_ll}
\end{figure}

The top plots in Figure \ref{fig_ll} show the BLF-scree plots obtained from the PARCOR approach
for each of the 50 datasets under
the two scenarios (i) $\phi_{1,1,2}=0$ and (ii) $\phi_{1,1,2}=-0.8$ for model orders
$m=1,\ldots,5.$ We see that in both scenarios 
the BLF-scree plots indicate that the optimal model order is $P=2.$ We also computed the DIC 
as explained in the previous section for each model order $m=1,\ldots,5$ and each dataset  
under scenarios (i) and (ii). The bottom left plot in Figure \ref{fig_ll} shows the distributions
of the optimal model orders chosen by the TV-VPARCOR and TV-VAR approaches for scenario (i), while the
right bottom plot shows the distribution for scenario (ii). We see that both, the TV-VPARCOR and TV-VAR
approaches lead to very similar results and model order 2 is adequately chosen as the 
optimal model order under the two scenarios for most of the 50 datasets.  

\begin{figure}[ht]
\centering
\includegraphics[width=5.0cm]{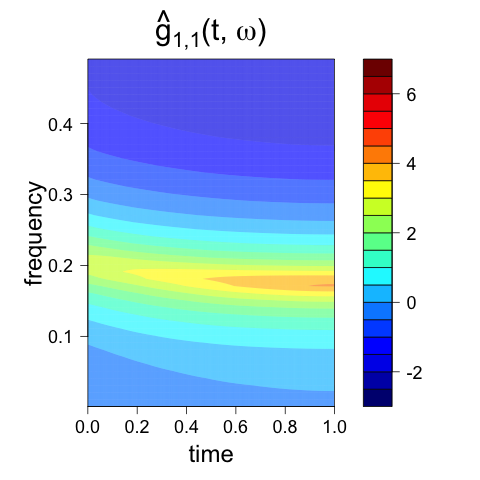}\hfill
\includegraphics[width=5.0cm]{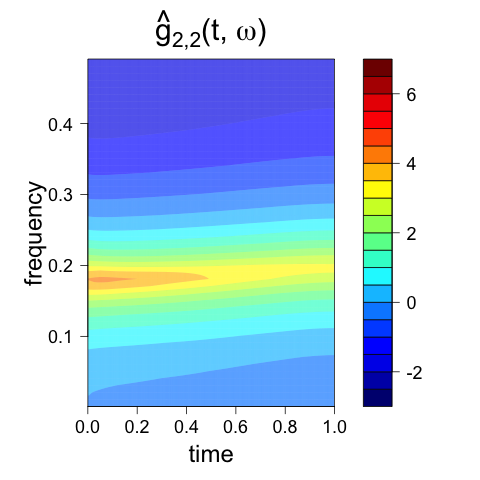}\hfill
\includegraphics[width=5.0cm]{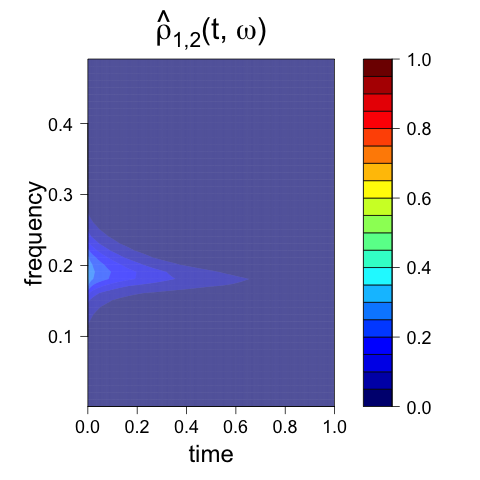}\hfill
	\caption{Case with $\phi_{1,1,2}=0.$ Left: Estimated average log spectral density of the first component. Middle: Estimated average log spectral density of the second component. Right: Estimated average squared coherence.}  \label{fig_est1}
\end{figure}

\begin{figure}[ht]
\centering
\includegraphics[width=5cm]{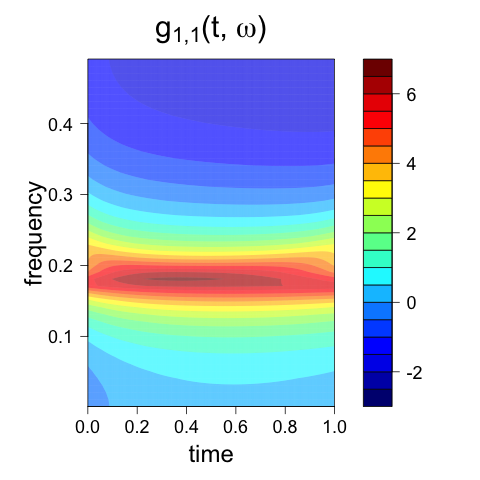}\hfill
\includegraphics[width=5cm]{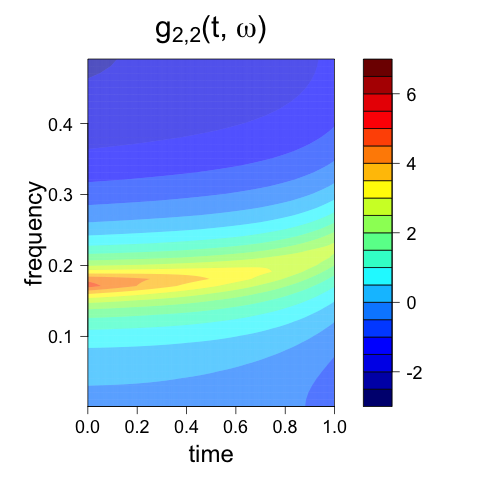}\hfill
\includegraphics[width=5cm]{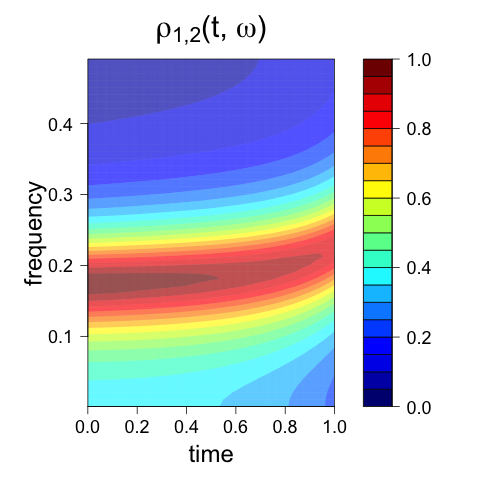}\hfill \\
\includegraphics[width=5cm]{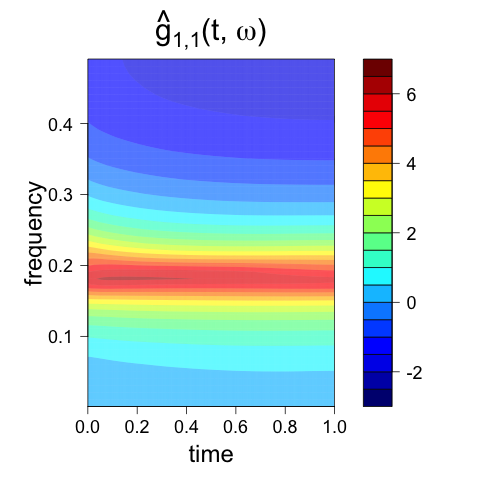}\hfill
\includegraphics[width=5cm]{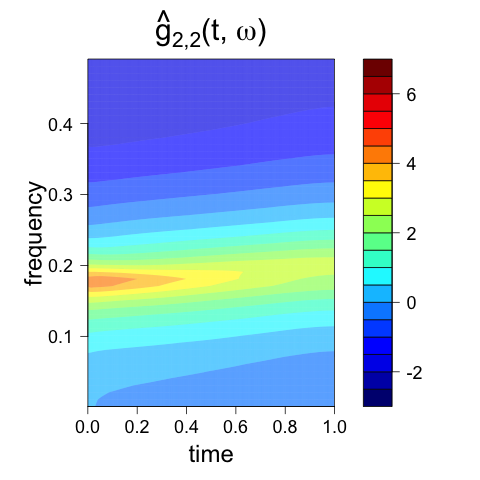}\hfill
\includegraphics[width=5cm]{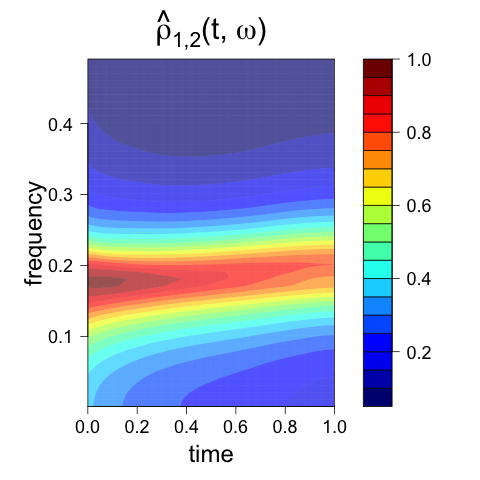}\hfill
\caption{Case with $\phi_{1,1,2}=-0.8$. Top: True log spectral density ${g}_{11}(t, \omega)$ (left), true log spectral density ${g}_{22}(t, \omega)$ (middle), true squared coherence $\rho^2_{1,2}(t,\omega)$ (right). Bottom: Estimated ${\hat{g}}_{11}(t, \omega)$ (left), estimated ${\hat{g}}_{22}(t, \omega)$ (middle), estimated $\hat{\rho}^2_{1,2}(\omega,t)$ (right). }   
\label{fig_ex3}
\end{figure}

Figures \ref{fig_est1} and \ref{fig_ex3} summarize posterior inference obtained from the TV-VPARCOR approach using a model order of 2 for the two scenarios (i) and (ii), respectively.
Estimated spectral densities were obtained from
the posterior means of the approximate smoothing distributions of the forward and backward PARCOR coefficient matrices over time.  The estimated log spectral densities displayed in the figures were obtained by averaging the estimated log spectral densities over the 50 simulated datasets. The bivariate TV-VPARCOR model is able to adequately capture the structure of the individual spectral densities and also that of the squared coherences. From these figures 
we also see that when $\phi_{1,1,2}=-0.8$, the second series has stronger impact on the first one and therefore their 
coherence is stronger.  The TV-VPARCOR model is able to adapt and adequately capture this feature. 

In order to compare the performance of the TV-VPARCOR and TV-VAR models in estimating the various time-frequency representations, we computed the mean and standard deviations of the average squared error (ASE) for each of the models in each of the two simulation scenarios. The ASE is defined as follows \citep{ombao2001automatic}
\begin{align}
    \mbox{ASE}_n &= (TL)^{-1}  \sum^T_{t=1}\sum^L_{l=1} \left(\log \hat{{\bm{g}}}(t, \omega_l) - \log {\bm{g}}(t, \omega_l)\right)^2,
\end{align}
where $\omega_l = 0, 0.001, 0.011, \dots, 0.5$. Note that we have $n=50$ simulated datasets for each of the two scenarios. Table \ref{table1} summarizes the mean and standard deviations of the ASE based on $\mbox{ASE}_n$ for scenarios (i) and (ii).  
Note that the simulated data are actually generated from TV-VAR$(2)$ models, not from TV-VPARCOR models, so we expect TV-VAR models to do better in terms of ASE for this specific simulation study. Nevertheless the
proposed TV-VPARCOR approach has comparable performance in terms of estimating the time-frequency characteristics of the original process while being computationally more efficient. 
\begin{table}[h]
	\begin{center}
\small
\begin{tabular*}{\hsize}{@{\extracolsep{\fill}}cccc}
\hline 
\\[-10pt]
	& \multicolumn{3}{c}{Case (i): $\phi_{1, 1, 2} = 0$} 
\\
$Model$ & ${g}_{11}$ & ${g}_{22}$ & $\rho^2_{12}$\\
PARCOR & $0.0246 (0.0183)$ & $0.0255 (0.0147)$ & $0.0008 (0.0006)$  \\
TV-VAR(2) & $0.0171 (0.0068) $ & $0.0186 (0.0080)$ & $0.0009 (0.0005)$\\
\hline
\\[-10pt]
	& \multicolumn{3}{c}{Case (ii): $\phi_{1, 1, 2} = -0.8$} \\
$Model$ & ${g}_{11}$ & ${g}_{22}$ & $\rho^2_{12}$\\
PARCOR & $0.0284 (0.0118)$ & $0.0238 (0.0086)$ & $0.0027 (0.0023)$ \\
TV-VAR(2)  & $0.0254(0.0073)$ & $0.0253 (0.0081)$ & $0.0023 (0.0011)$\\
\hline
\end{tabular*}
	\end{center}
	\caption{Mean ASE values and corresponding standard deviations (in parentheses) obtained
	from TV-VPARCOR and TV-VAR$(2)$ models for 
	the TV-VAR$(2)$ simulated data.}\label{table1}
\end{table}
In fact, Table \ref{table2} presents the computation times
for both models again averaging over the 50 realizations in each case. We see that even for this example with only two time series
components and a model order of 2, the TV-VPARCOR models require almost half of the computation time required by the 
TV-VAR$(2)$ models. As the model order and the number of time series components increase, differences
in computational will be more pronounced, making the TV-VPARCOR approach more efficient for modeling 
large temporal datasets. 
\begin{table}
\small
	\begin{center}
\begin{tabular*}{\hsize}{@{\extracolsep{\fill}}ccc}
\hline
\\[-10pt]
	Model & (i): $\phi_{1, 1, 2} = 0$  & (ii): $\phi_{1,1, 2} = -0.8$ \\
\hline
\\[-10pt]
PARCOR & $558s$  & $521s$  \\
TV-VAR & $925s $ & $828s$\\ \hline
\end{tabular*}
	\end{center}
	\caption{Computation times (in seconds) for TV-VPARCOR and TV-VAR models.}\label{table2}
\end{table}


\subsection{20-dimensional TV-VAR(1)} 

We analyze data simulated from a 20-dimensional non-stationary TV-VAR$(1)$ process with $T=300$ in which the $(i,j)$ elements of the matrix of VAR coefficients at time $t,$ $\bm{\Phi}_{t},$ are given as follows:
\begin{eqnarray*}
\bm{\Phi}_{t}(i,j) & = & \left\{ 
\begin{array}{cll}
0.7 + \frac{0.2}{299} \times t & \mbox{for all} & i=j, \;\; i=1,\ldots,10,  \\
-0.95 + \frac{0.2}{299}\times t & \mbox{for all} & i=j, \;\; i=11,\ldots,20, \\
0.9 & \mbox{for} & (i,j) \in \{(1,5), (2,15)\},  \\
-0.9 & \mbox{for} & (i,j) \in \{(6,12), (15,20)\},  \\
0 & \mbox{otherwise}.
\end{array}
\right. 
\end{eqnarray*}
for $t=1,\ldots,300.$ In addition, we assume $\bm{\Sigma}=0.1 \bm{I}_{20}.$
  
We fit TV-VPARCOR models considering $P_{max} = 3.$ Note that the PARCOR approach with $P_{max}=3$ requires 
fitting 6 multivariate DLMs with state-space parameter vectors of dimension $400.$ Alternatively,
working directly with TV-VAR representations with $P_{max}=3$ requires fitting 3 multivariate DLMs with
state-space parameter vectors of dimension $400$ for model order 1, $800$ for model order 2, and $1200$ for model order 3. The TV-VAR model representation leads to a rapid increase of the dimension of the state-space vector with the model order, which significantly reduces the computational efficiency, particularly for large and even moderate $T.$ The TV-VPARCOR approach requires fitting more multivariate DLMs, but the dimensionality of the state-space vectors remains constant with the model order. This is an important advantage of the TV-VPARCOR approach. 
In fact, the TV-VPARCOR model required 585s of computation time 
for $P_{\max}=3,$ while the TV-VAR model required 3379s with the same $P_{\max}=3$ value. Posterior computations were completed in both cases using a MacBookPro13 with Intel Core i5, with a
2 GHz (1 Processor). Note also that, for a given model order the PARCOR approach can be further optimized 
in terms of computational efficiency, as the forward and backward DLMs can run in parallel. 

We assumed  prior hyperparameters $\bm{m}_{0,\cdot, m} = \bm{0}$ and $\bm{C}_{0, \cdot, m} = \bm{I}_{400}$ for the forward and backward PARCOR models. The elements of the diagonal component of discount factor matrices, $\delta_{f,m}$ and $\delta_{b,m},$ were chosen from a grid of values in $(0.99, 1)$. As mentioned above we also fit TV-VAR models with model orders going from $1$ to $3$ using similar prior hyperparameters and discount factors. For both types of models the DIC picked model order $1$ as the optimal model order, which is the  corresponding true model order in this case. Both types of models led to similar posterior inference of the time-frequency spectra.

\begin{figure}
    \centering
    \includegraphics[width=3.5cm]{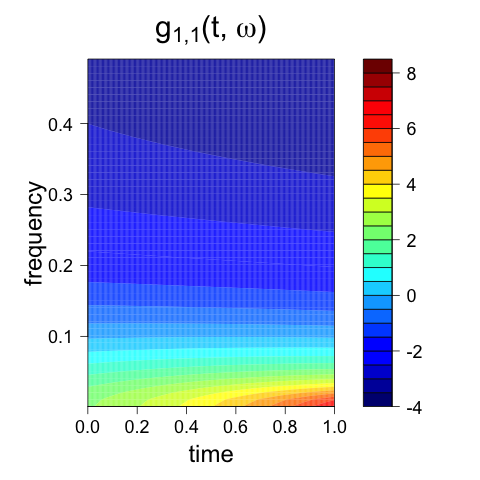}\hfill 
    \includegraphics[width=3.5cm]{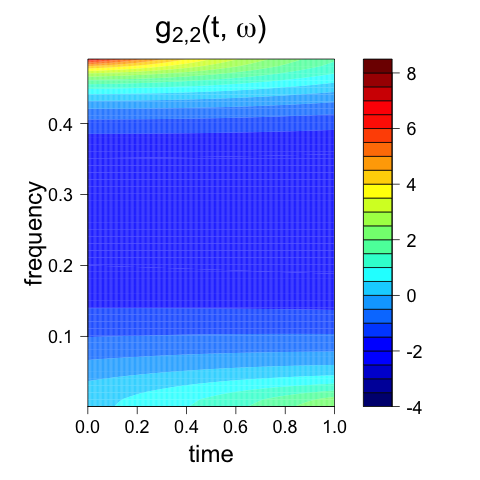}\hfill 
    \includegraphics[width=3.5cm]{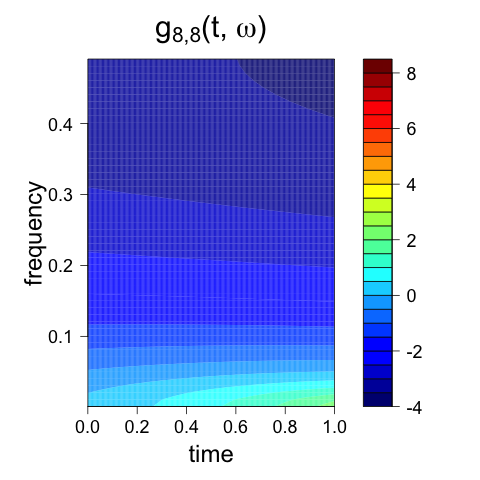}\hfill 
    \includegraphics[width=3.5cm]{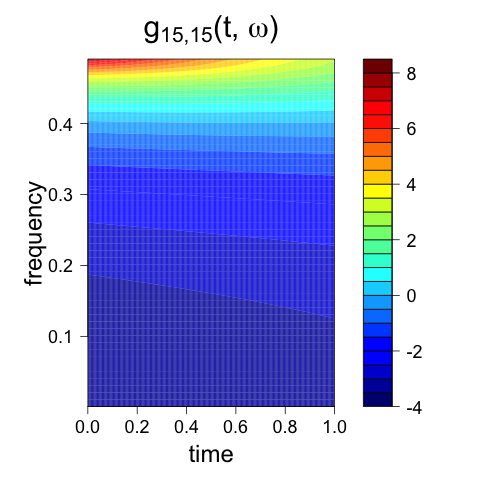}\hfill \\
    \includegraphics[width=3.5cm]{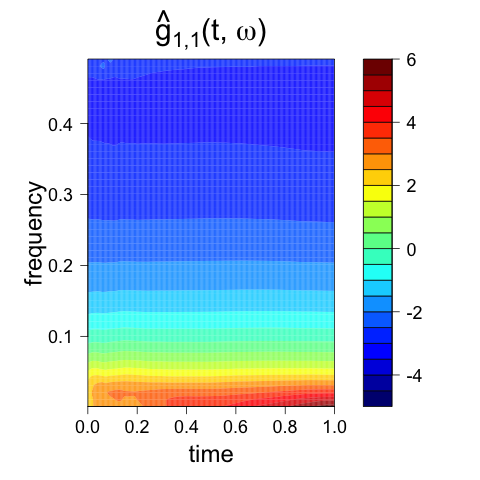}\hfill 
    \includegraphics[width=3.5cm]{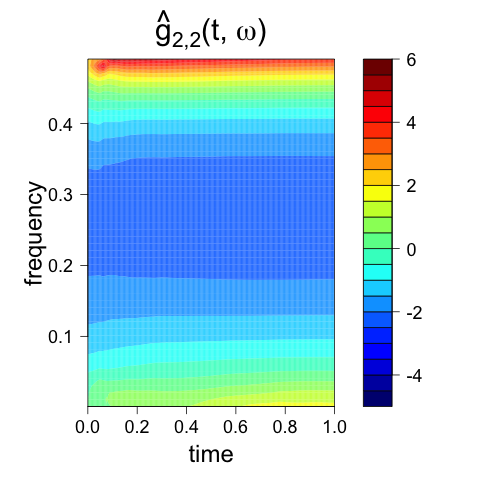}\hfill 
    \includegraphics[width=3.5cm]{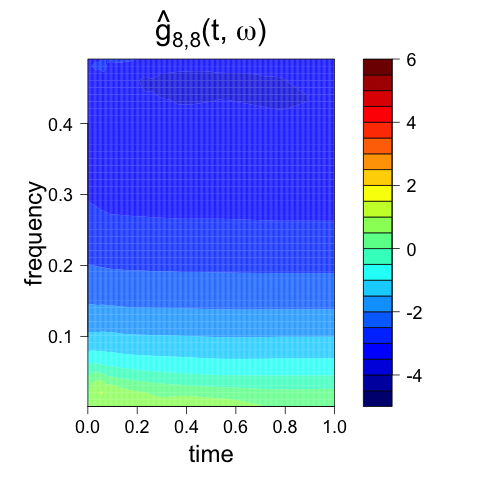}\hfill 
    \includegraphics[width=3.5cm]{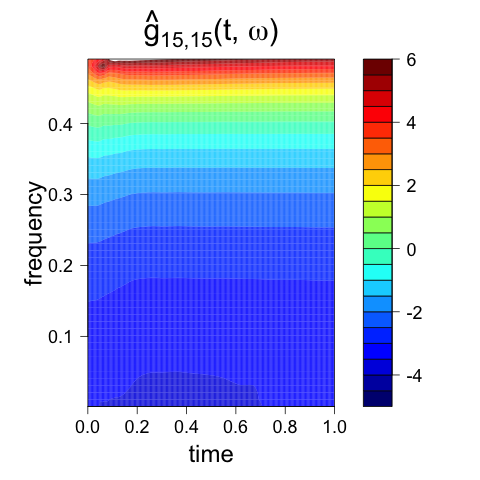}\hfill 
    \caption{Top: True log spectral densities of time series components 1, 2, 8 and 15. Bottom: estimated log spectral densities of the same components obtained from the PARCOR approach with model order 1. }
    \label{VAR(20)_sd}
\end{figure}

\begin{figure}
    \centering
    \includegraphics[width=3.5cm]{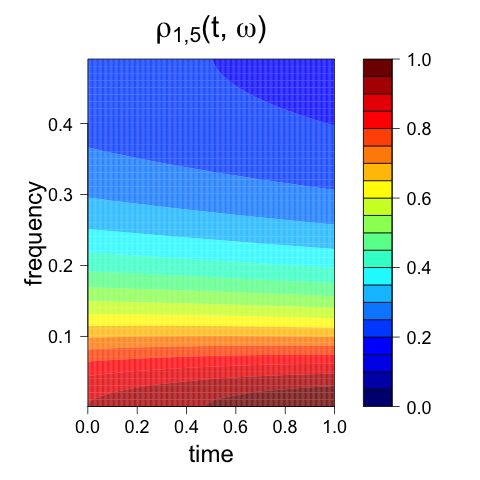}\hfill 
    \includegraphics[width=3.5cm]{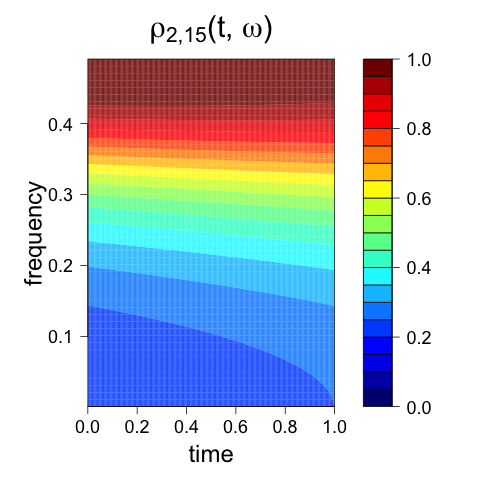}\hfill 
    \includegraphics[width=3.5cm]{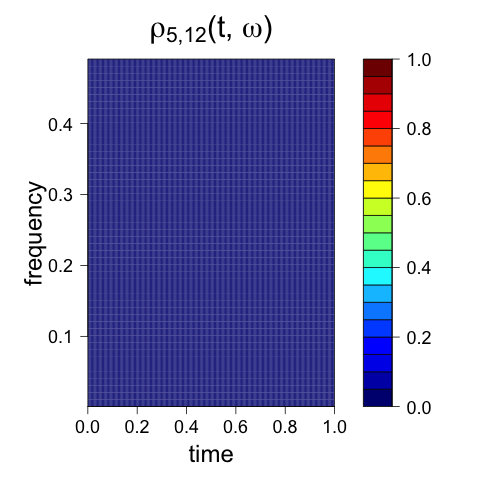}\hfill 
    \includegraphics[width=3.5cm]{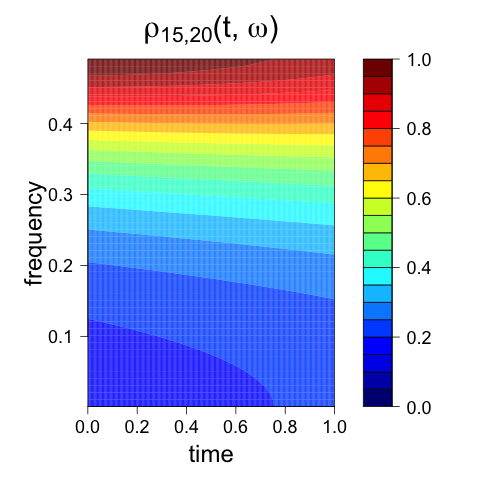}\hfill \\
    \includegraphics[width=3.5cm]{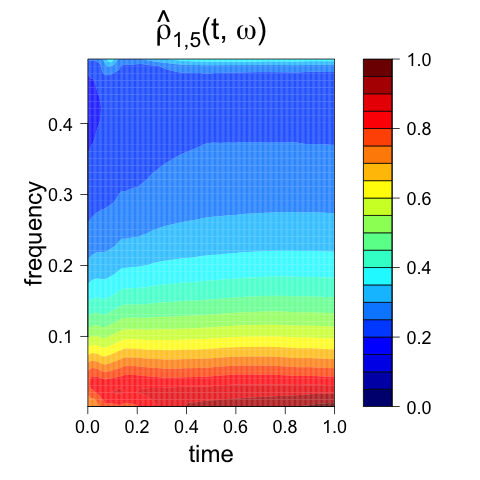}\hfill 
    \includegraphics[width=3.5cm]{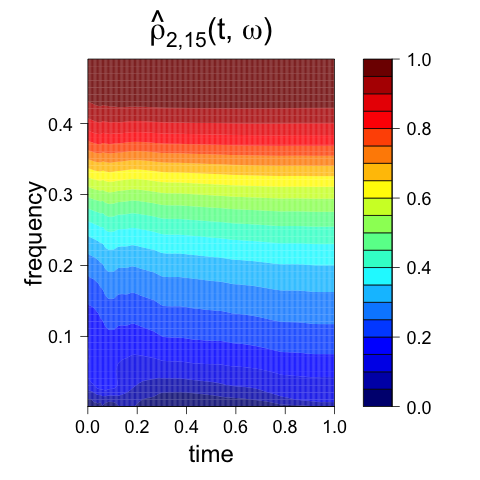}\hfill 
    \includegraphics[width=3.5cm]{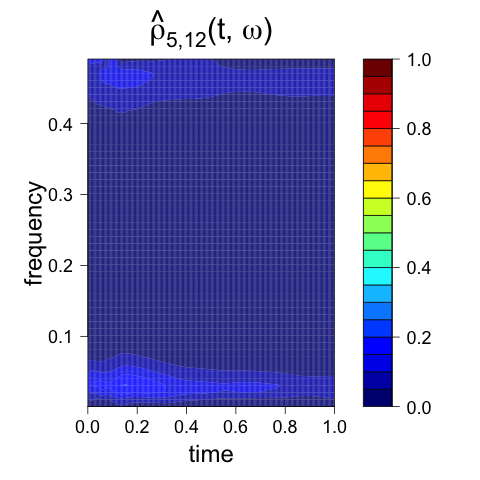}\hfill 
    \includegraphics[width=3.5cm]{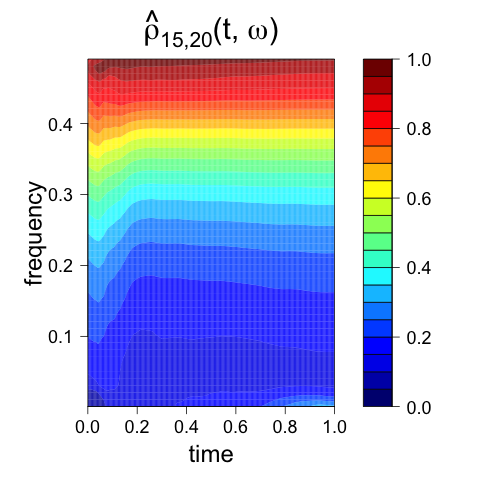}\hfill 
    \caption{Top: True coherence between components 1 and 5, 2 and 15, 5 and 12, and 15 and 20. Bottom: Corresponding estimated coherences obtained from the PARCOR model. }
    \label{VAR(20)_PARCOR_coh}
\end{figure}

Here we only show the results from the TV-VPARCOR approach. 
Figure \ref{VAR(20)_sd} shows the true and estimated log spectral densities from the TV-VPARCOR model for 4 components of the 20-dimensional time series, namely, components $1$, $2$, $8$ and $15$. Figure \ref{VAR(20)_PARCOR_coh} shows the true and estimated coherences between components $1$ and $5$, components $2$ and $15$, components $5$ and $12$, and components $15$ and $20$. Overall we see that the TV-VPARCOR approach adequately captures the space-time characteristics of the original multivariate non-stationary time series process. Furthermore, the TV-VPARCOR approach led to similar posterior estimates of the VAR coefficients
over time to those obtained from using a DLM representation of a TV-VAR (see Figure 1 in the Supplementary Material).

\section{Case studies}
\subsection{Analysis of multi-channel EEG data}
\begin{figure}[htp]
\centering
\includegraphics[width = 5cm]{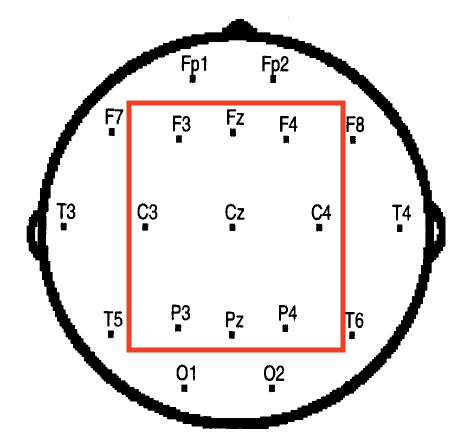}\hfill
\caption{Representation of the \code{lctal19} electrode placement. Here we focus on the nine channels in the region highlighted.} \label{fig_rep}
\end{figure}

We consider the analysis of multi-channel EEG data recorded on a patient that received electroconvulsive therapy (ECT) as a treatment for major depression. These data are part of a larger dataset, code named \code{Ictal19}, that corresponds to recordings of $19$ EEG channels from one subject during ECT. As an illustration, we use our multivariate TV-VPARCOR model to analyze $9$ of the channels, specifically channels $F_3, F_z, F_4, C_3, C_z$, $C_4, P_3, P_z, P_4$ shown in Figure \ref{fig_rep}. We chose these channels because the are closely located and because based on previous analyses we expect strong similarities in their temporal structure over time. The full multi-channel dataset was analyzed in \cite{west1999evaluation} and \cite{prado2001multichannel} using univariate TVARs separately for each channel, and also using dynamic regression models. The original recordings of about $26,000$ observations per channel were subsampled every sixth observation from the highest amplitude portion of the seizure, leading to a set of series of $3,600$ observations (corresponding to $83.72s$) per channel \citep{prado2001multichannel}.

We analyzed the $K=9$ series listed above jointly using a multivariate TV-VPARCOR model. We
considered a maximum model order of $P_{\max}=20$ and discount factor values on a grid in the $(0.99,1]$ range (with equal spacing of $0.001$).
We further assumed that the discount factor values were the same across channels. This assumption was based on previous analyses of the individual channels using univariate TVAR models that showed similar optimal discount values for the different channels. We set $n_{0, f, m} = n_{0, b, m} = 1,$ and $S_{0, f, m} = S_{0, b, m} = 2000 \bm{I}_9$ for all $m.$ In addition, we set the same initial prior parameters $\bm{m}_{0, f, m} = \bm{m}_{0, b, m} = \bm{0}$ and $\bm{C}_{0, f, m} = \bm{C}_{0, b, m} = 1000\bm{I}_{81}.$ 
The optimal model order was found to be 5 (see Figure \ref{EEG_DIC} in the Supplementary Material) and so, the results presented here correspond to a TV-VPARCOR model with this order. Higher order models were also fitted leading to similar but slightly smoother results in terms of the estimated spectral density, coherence and partial coherence.  

\begin{figure}
\centering
\includegraphics[width=5cm]{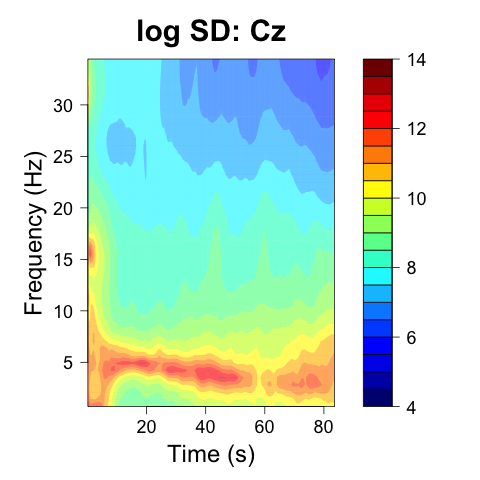}\hfill
\includegraphics[width=5cm]{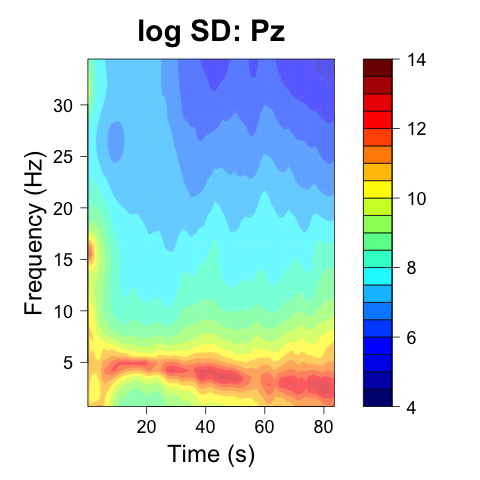}\hfill
\includegraphics[width=5cm]{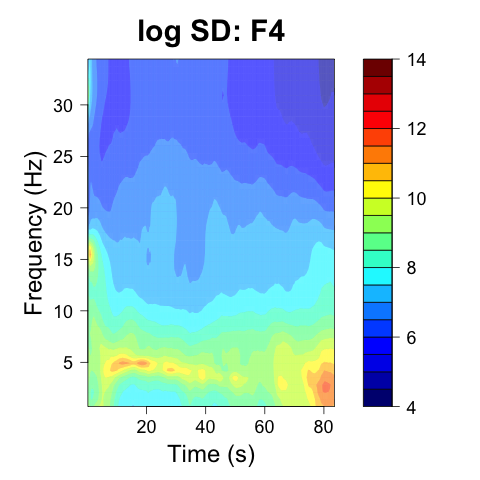}\hfill
\caption{Estimated log-spectral densities for channels Cz, Pz and F4.}
\label{EEG_SD}
\end{figure}

Figure \ref{EEG_SD} displays estimated log spectral densities of channels Cz, Pz and F4. We note that the multi-channel EEG data are dominated by frequency components in the lower frequency band (below 18Hz). Furthermore, each EEG channel shows a decrease in of the dominant frequency over time, starting around 5Hz and ending around approximately 3Hz. This decrease in the dominant frequency was also found in \cite{west1999evaluation}. Channels Cz and Pz are 
more similar to each other than to channel F4 in terms of their log-spectral densities. The three channels show the largest power around the same frequencies, however, channel F4 displays smaller values in the power log-spectra than those for channels Cz and Pz. The remaining channels also show similarities in their spectral content (not shown).

\begin{figure}[ht]
    \centering
    \includegraphics[width = 5cm]{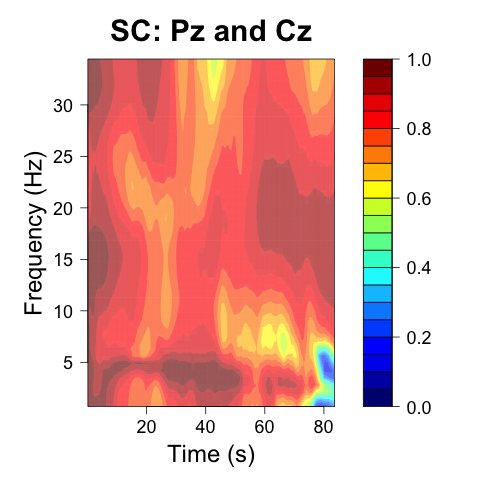}\hfill
    \includegraphics[width = 5cm]{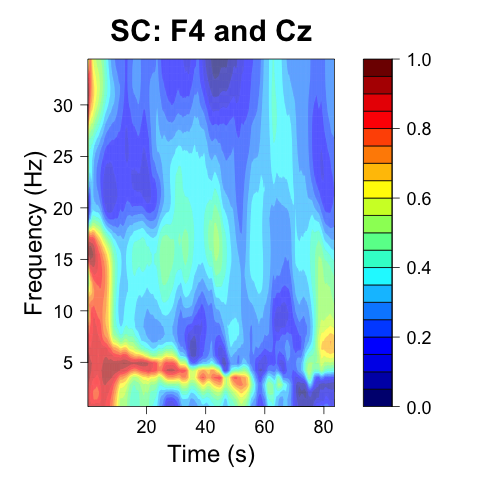}\hfill
    \includegraphics[width = 5cm]{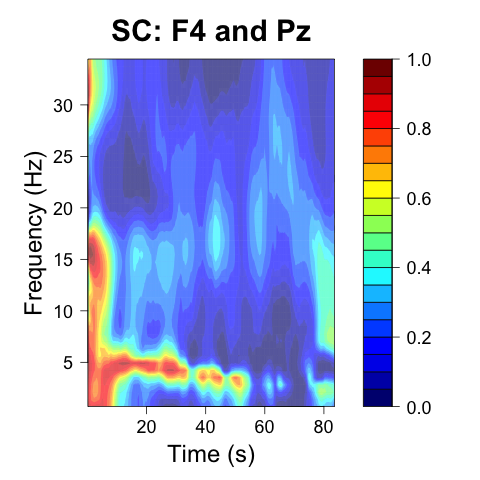}\hfill\\
    \includegraphics[width = 5cm]{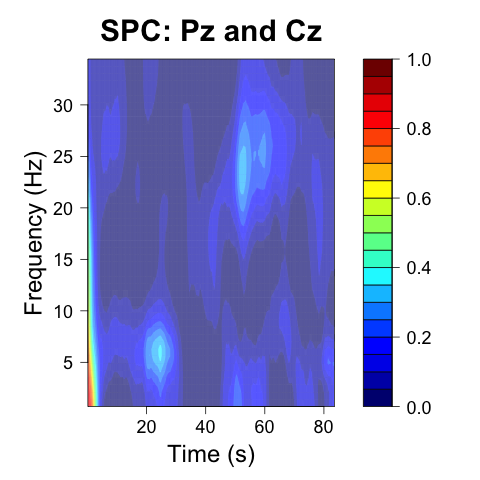}\hfill
    \includegraphics[width = 5cm]{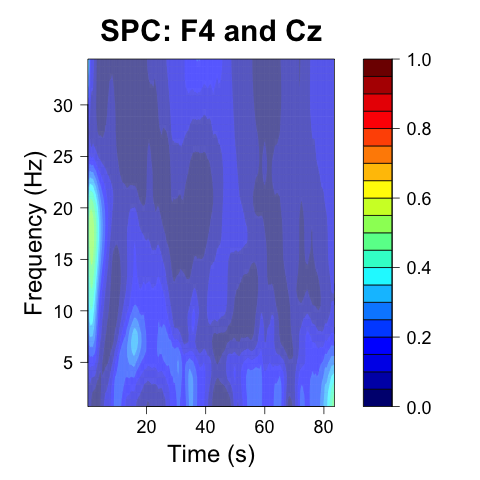}\hfill
    \includegraphics[width = 5cm]{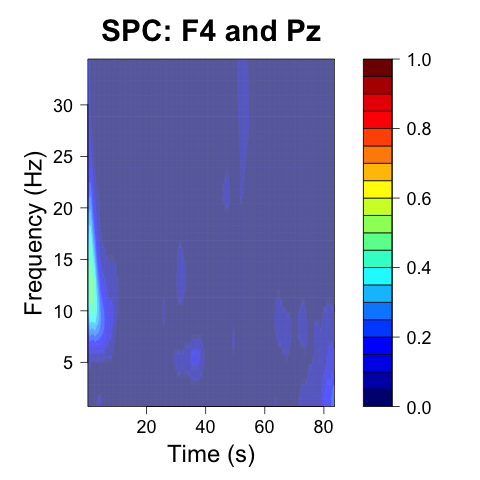}\hfill
    \caption{Top plots: Squared coherence between Pz and Cz, F4 and Cz, and F4 and Pz, respectively. Bottom plots: Squared partial coherence between Pz and Cz, F4 and Cz, and F4 and Pz.}
    \label{EEG_SC}
\end{figure}

Figure \ref{EEG_SC} shows estimated squared coherences (top) and estimated squared partial coherences (bottom) between channels Pz and Cz, F4 and Cz, and F4 and Pz. Channels Pz and Cz show a very strong coherence over time across almost all the frequency bands under 35Hz. On the other hand, channel F4 shows strong coherence with channels Pz and Cz across frequencies below 15-18 Hz at the beginning of the seizure. After the initial 10-15s, and approximately until about 50s, there is a strong coherence between F4 and Pz and Cz only at the dominant frequency of 3-5Hz that dissipates towards the end of the seizure.
The partial coherence across pairs of channels is the frequency domain version of the squared correlation coefficient between relationship between pairs of components after the removal of the effects of all the other components. Figure \ref{EEG_SC} shows that the estimated squared partial coherences between Pz and Cz, F4 and Cz and F4 and Pz are essentially negligible for most frequency bands over the seizure course. This makes sense due to the  fact that most of the 9 EEG channels are so strongly coherent across different frequency bands over the entire period of recording. The estimated squared partial coherence between channels Pz and Cz is large for frequencies below 5Hz only at the very beginning of the seizure. These findings are consistent with results from the analysis of these data in \cite{west1999evaluation} and \cite{prado2001multichannel}.

\subsection{Analysis of multi-location wind data} 

We analyze wind component data derived from median wind speed and direction measurements taken every 4 hours from June 1st 2010 to August 15th in 3 stations in Northern California.  These data were obtained from the Iowa Environmental Mesonet (IEM) Automated Surface Observing System (ASOS) Network, a publicly available database 
(see {\tt http://mesonet.agron.iastate.edu/ASOS/}). ASOS stations are located at airports and take observations and basic reports from the National Weather Service (NWS), the Federal Aviation Administration (FAA), and the Department of Defense (DOD). For additional information about the ASOS measurements see \cite{ASOSmanual}. Here we analyze time series data from Monterey, Salinas and Watsonville, 3 stations located near the Monterey Bay.

We use the TV-VPARCOR approach for joint analysis of the six-dimensional time series corresponding to the wind time series components for the 3 stations. We set $P_{max} = 10$ and consider discount factor values on a grid in the $(0.9,1]$ range.  We assume that discount factor values were the same across components for the 3 stations. We set the prior hyperparameters as follows:  $n_{0, f, m} = n_{0, b, m} = 1,$ and $S_{0, f, m} = S_{0, b, m} = 5\bm{I}_6,$  $\bm{m}_{0, f, m} = \bm{m}_{0, b, m}= \bm{0}$  and $\bm{C}_{0, f, m} = \bm{C}_{0, b, m} = 10\bm{I}_{36}$ for all $m.$
The optimal model order chosen by the approximate DIC calculation is $P=3$ (see Figure \ref{screeplot_wind} in the Supplementary Material). For this model order we found that the optimal discount factors were $0.97, 0.97$ and $0.99,$ respectively, for each of the 3 levels of the forward PARCOR model, and $0.98,$ $0.98$ and $0.99$ for each of the 3 levels of the backward PARCOR model. 

\begin{figure}
    \centering
    \includegraphics[width = 5cm]{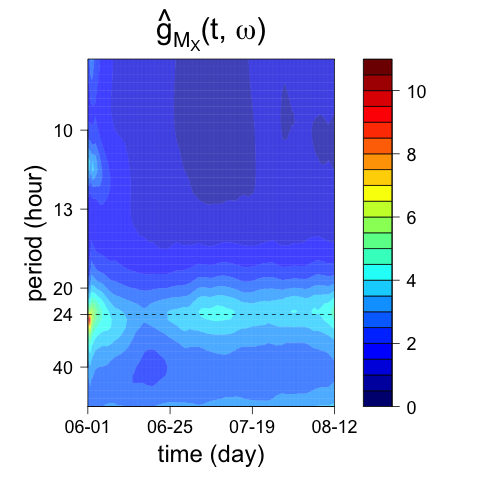}\hfill
    \includegraphics[width = 5cm]{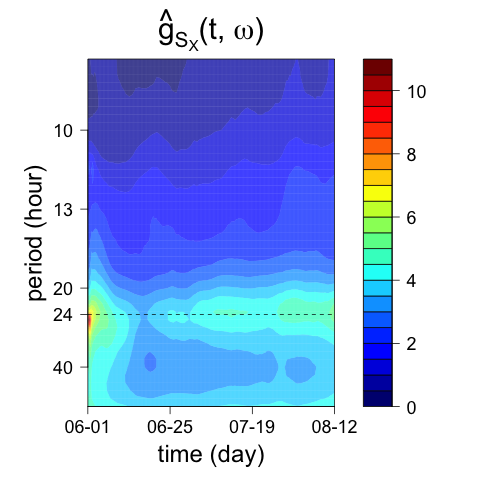}\hfill
    \includegraphics[width = 5cm]{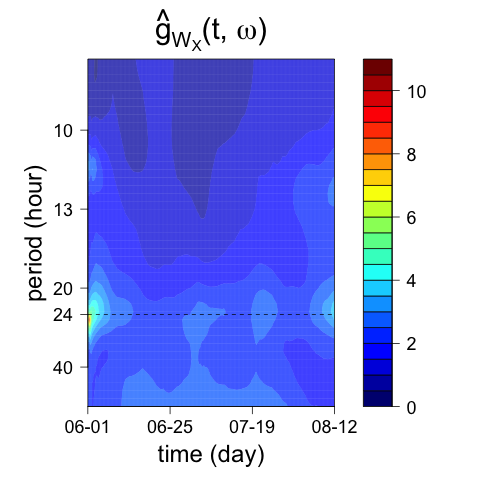}\hfill \\
    \includegraphics[width = 5cm]{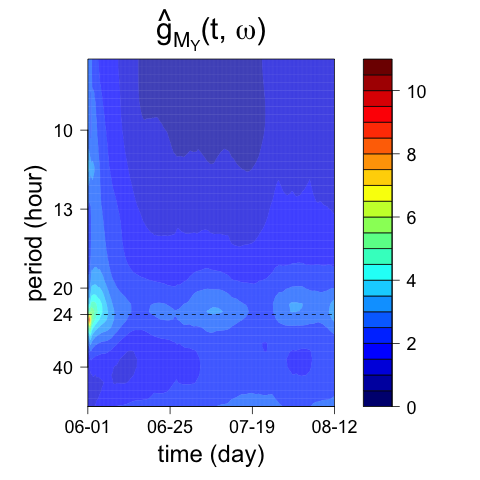}\hfill
    \includegraphics[width = 5cm]{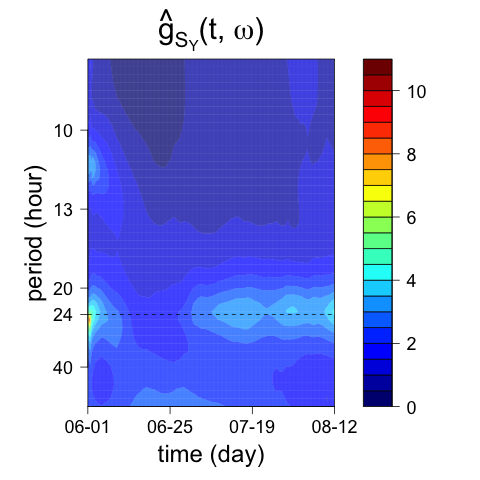}\hfill
    \includegraphics[width = 5cm]{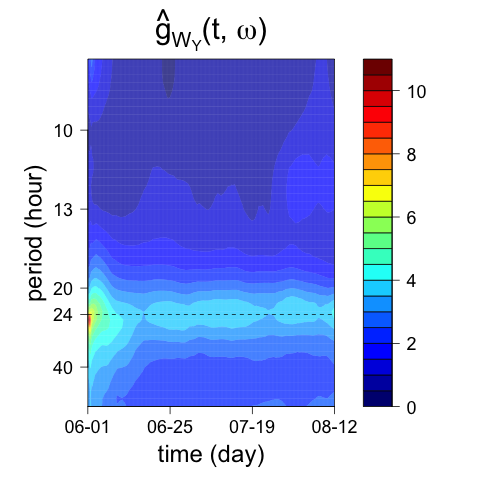}\hfill 
    \caption{Top row: Estimated log-spectral densities of the East-West (X) components for Monterey, Salinas and Watsonville. Middle row: Estimated log-spectral densities of the North-South (Y) components for Monterey, Salinas and Watsonville.}
    \label{wind_sd}
\end{figure}

Figure $\ref{wind_sd}$ shows the estimated log spectral densities of the East-West component (X component) and the North-South component (Y component) for each location. We can observe that there is a dominant quasi-periodic behavior around the 24 hour period for the East-West (X) components in Monterey and Salinas, as well as the North-South (Y) component in Watsonville.  This quasi-periodic behavior is also present, although is less persistent over time, in the East-West component in Watsonville and the North-South components in Monterey and Salinas.  The observed quasi-periodic pattern observed in the estimated log-spectral for these three locations is consistent with the fact that stronger winds are usually observed in the afternoons/evenings during the summer in these locations, while calmer winds are observed during the rest of the day. Note also that the quasi-periodic daily behavior is more persistent over the entire set of summer months for the North-South component than the East-West component in Watsonville, while the quasi-periodic behavior is more persistent in the East-West component than in the North-South component in 
Monterey and Salinas.

\color{black}

\begin{figure}
    \centering
    \includegraphics[width = 5cm]{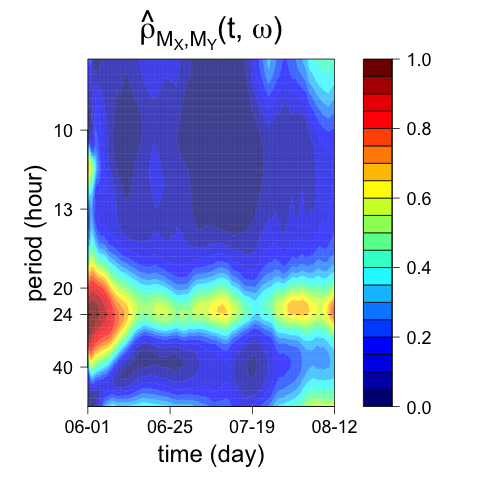}\hfill
    \includegraphics[width = 5cm]{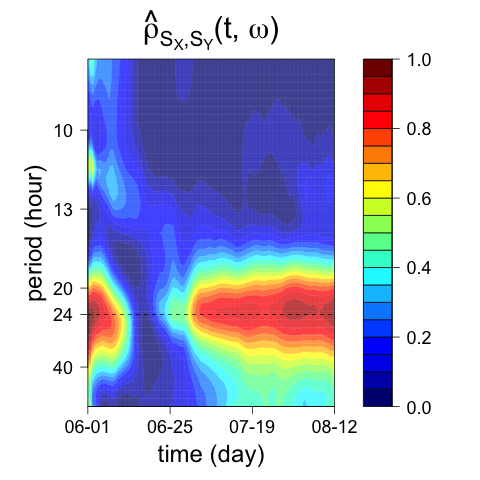}\hfill
    \includegraphics[width = 5cm]{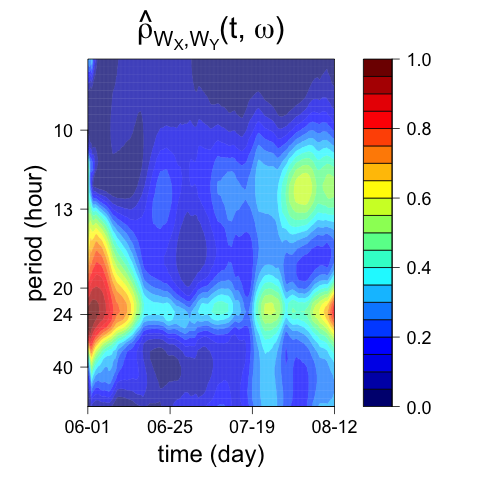}\hfill\\
    \includegraphics[width = 5cm]{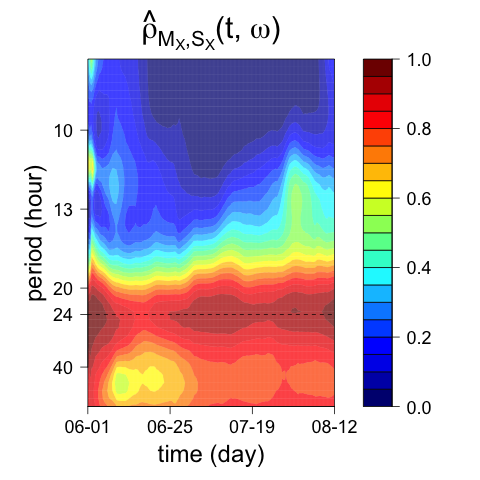}\hfill
    \includegraphics[width = 5cm]{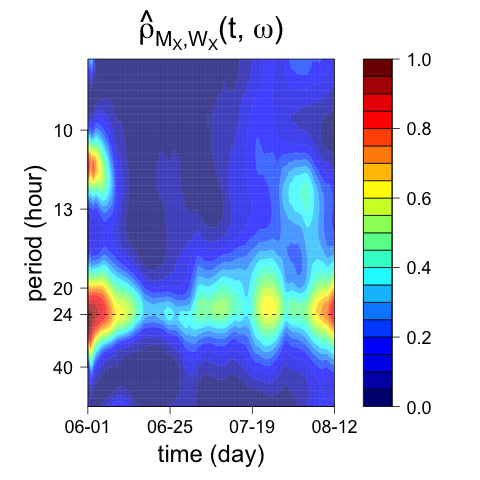}\hfill
    \includegraphics[width = 5cm]{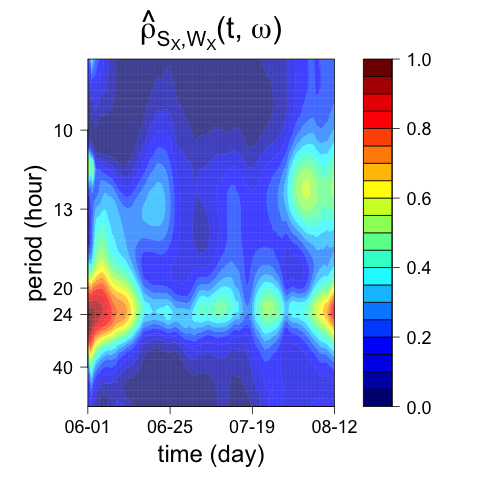}\hfill\\
    \includegraphics[width = 5cm]{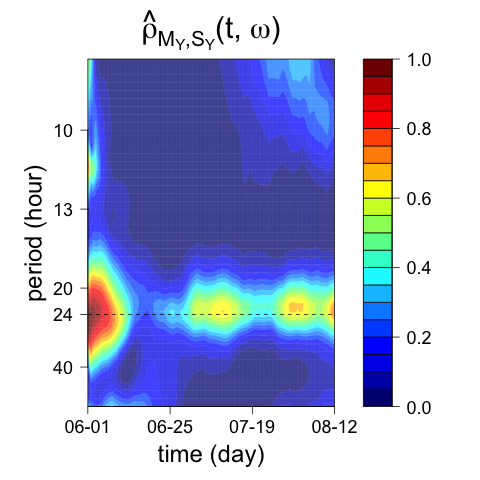}\hfill
    \includegraphics[width = 5cm]{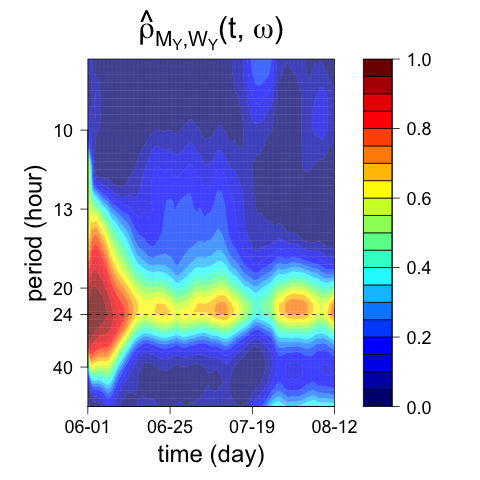}\hfill
    \includegraphics[width = 5cm]{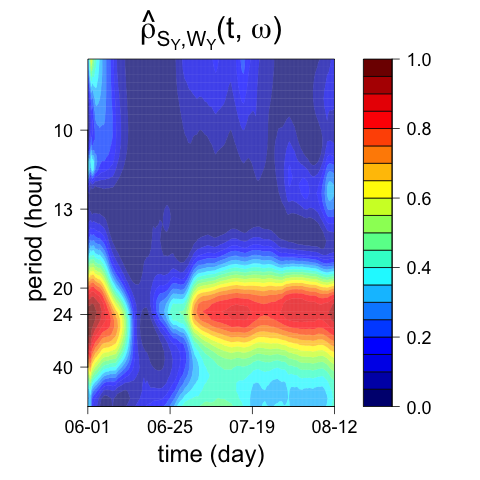}\hfill
    \caption{Top row: Estimated squared coherences between the East-West (X) component and North-South (Y) component in Monterey, Salinas and Watsonville. Middle row: Estimated squared coherences between the East-West (X) components of Monterey and Salinas, Monterey and Watsonville, and Salinas and Watsonville. Bottom row: Estimated squared coherences between 
    the North-South components in Monterey and Salinas, Monterey and Watsonville, and Salinas and Watsonville.}
    \label{wind_coh}
\end{figure}

Figure \ref{wind_coh} shows the estimated squared coherences  between each pair of wind components across the  three locations. There is a very strong coherence between Monterey and Salinas in the East-West (X) components for periods above 15 hours, with the strongest  relationship observed around 24 hours. 
We also observe that in general, there is a strong coherence between all the components around the 24 hours period. This coherence relationship tends to be more marked across some locations during the month of June (e.g., between the North-South components of Monterey and Salinas).
Furthermore, the estimated squared partial coherence (see Figure \ref{wind_pcoh} in the Supplementary Material) between the East-West components of Monterey and Salinas also shows that there is a relatively large linear relationship between these components for periods above 13 hours even after removing of the effect of all the other components for these locations and also after removing the effect of the wind components in  Watsonville.

\begin{figure}
    \centering
     \includegraphics[width = 14cm]{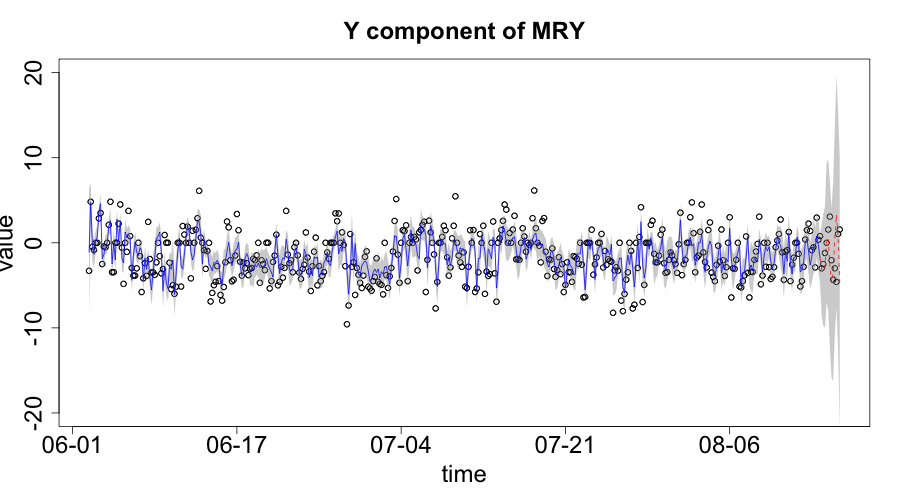} 
    \caption{Observed Monterey North-South Wind Component (dots); smoothed estimates 
	obtained from the posterior mean values of the TV-VPARCOR model (solid red line) and corresponding 90\% bands (gray shade); 72 hours forecast (dotted red line) and corresponding 90\% bands (gray shade).}
	\label{fig:forecast_wind}
\end{figure}

The TV-VPARCOR model can also be used for forecasting as described in Section 2.5. Figure \ref{fig:forecast_wind} shows 72 hours forecasts obtained from the TV-VPARCOR model for the North-South wind component in Monterey. We see that the model adequately captures the general future behavior of this time series component. 

\section{Discussion}
We consider a computationally efficient approach for analysis and forecasting of
non-stationary multivariate time series. We propose a multivariate dynamic linear modeling framework to describe the evolution of the PARCOR coefficients of a multivariate time series process over time. We use approximations in 
this multivariate TV-VPARCOR setting to obtain 
computationally efficient and stable inference and forecasting in the time and time-frequency domains.  The approximate posterior distributions derived from our approach are all of standard form. We also provide a method to choose the optimal number of stages in the TV-VPARCOR model based on an approximate DIC calculation. In addition, our 
model can provide reliable short term forecasting. 

The proposed framework provides computational efficiency and excellent performance in terms of the average squared error between the true and estimated time-varying spectral densities as shown in simulation studies and in the analysis of two multivariate time series datasets. The TV-VPARCOR
model representations also lead to very significant reduction in computational time when compared to 
TV-VAR model representations, particularly for cases in 
which we have model orders larger than 2-3 and more than
a handful of time series components.

In addition to simulation studies we have shown that the TV-VPARCOR approach can be successfully used
to analyze real multivariate non-stationary time series data. We presented the analysis of non-stationary multi-channel EEG data and also the analysis and forecasting of multi-location wind data. In the EEG case, our model was able to adequately detect the main time-frequency characteristics of individual EEG channels as well as the relationships across multiple channels over time. For the multi-location wind component data, our model detected a quasi-periodic pattern through the estimated spectral densities of each time series component which is consistent with the expected behavior of these components during the summer for locations near the Monterey Bay area. The model was also able to describe the time-varying relationships across multiple components and locations and led to reasonable short term forecasting.

The proposed dynamic multivariate PARCOR approach is computationally efficient when compared to state-space representations TV-VAR 
models. However, in many practical settings we may expect sparsity in the model parameters or situations in which
some parameters change over time and others do not. 
Future work will explore inducing, possibly time-varying, sparsity and dimension reduction in these multivariate TV-VPARCOR models while maintaining computational efficiency and accuracy in inference and forecasting.

\newpage 
\bibliographystyle{plainnat}
\bibliography{parcor}

\newpage
\section*{Supplementary Material}

\setcounter{figure}{0}
\subsection*{Supplementary figures for the 20-dimensional TV-VAR(1) example} 
\begin{figure}[h]
    \centering
    \includegraphics[width = 6cm]{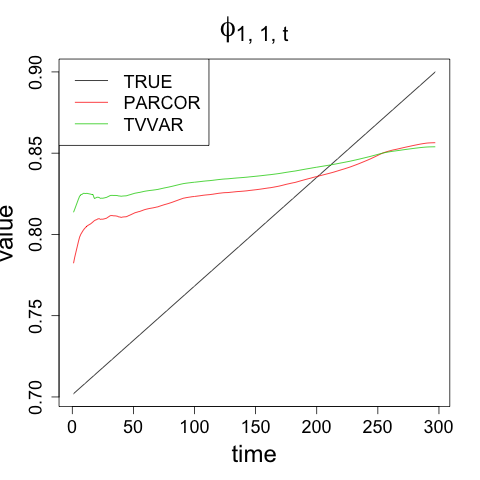}\hfill
    \includegraphics[width = 6cm]{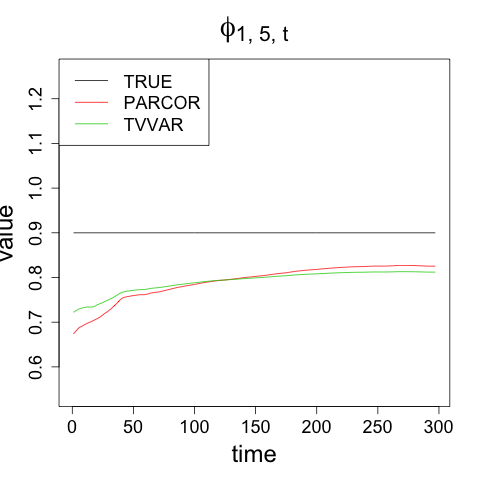}\hfill\\
    \includegraphics[width = 6cm]{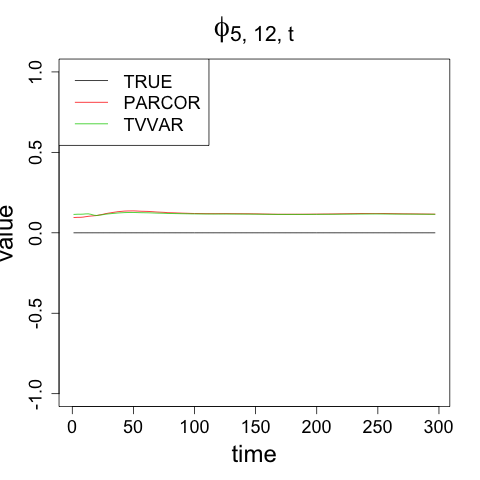}\hfill
    \includegraphics[width = 6cm]{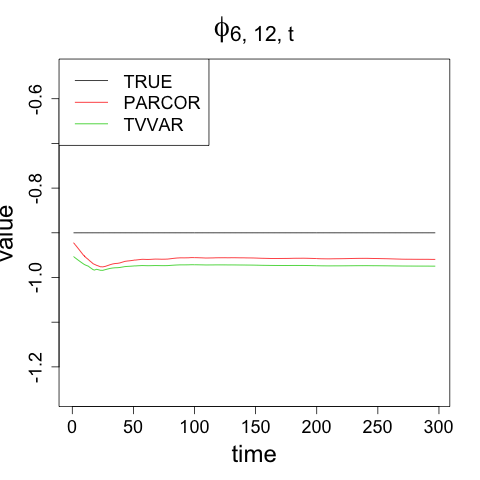}\hfill
    \caption{True and estimate traces of the TV-VAR coefficients $\phi_{1,1,t},$ $\phi_{1,5,t},$ $\phi_{6,12,t}$ and $\phi_{5,12,t}$ obtained from the TV-VPARCOR and TV-VAR approaches.}
    \label{fig:supplementary_1}
\end{figure}

\newpage
\subsection*{Supplementary figures for the multi-channel EEG analysis} 
\begin{figure}[h]
    \centering
    \includegraphics[width = 7cm]{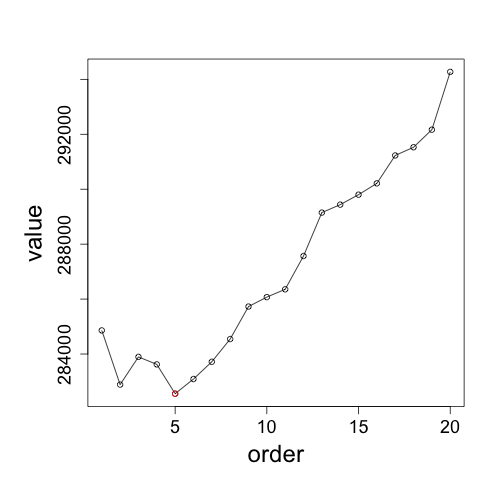}
    \caption{Multi-channel EEG data: DIC values.}
    \label{EEG_DIC}
\end{figure}

\newpage
\subsection*{Supplementary figures for the multi-location wind component data} 
\begin{figure}[h]
    \centering
    \includegraphics[width = 7cm]{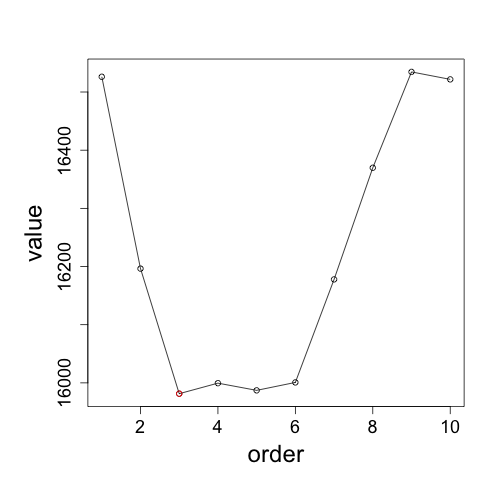}
    \caption{Multi-location wind components: DIC values.}
    \label{screeplot_wind}
\end{figure}

\begin{figure}
    \centering
    \includegraphics[width = 5cm]{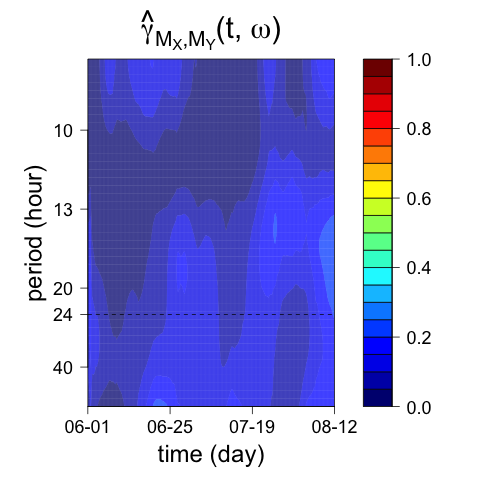}\hfill
    \includegraphics[width = 5cm]{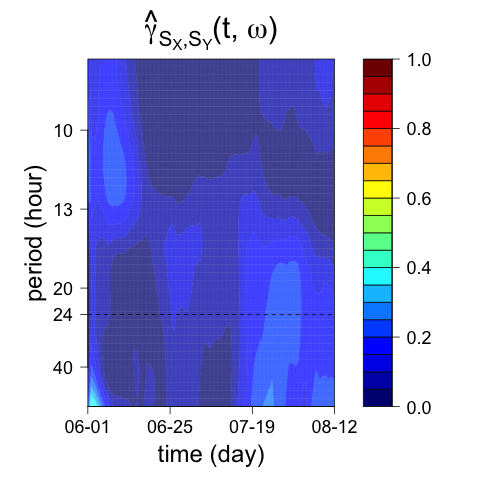}\hfill
    \includegraphics[width = 5cm]{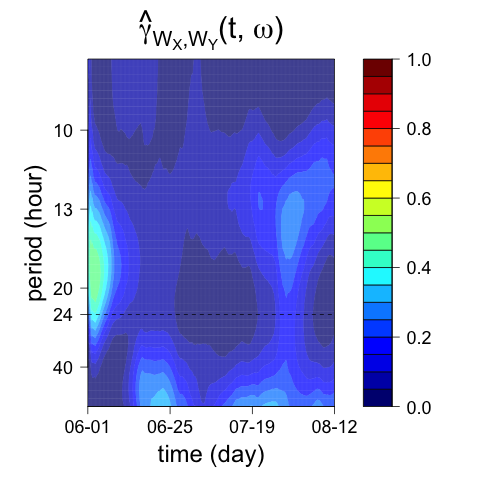}\hfill\\
    \includegraphics[width = 5cm]{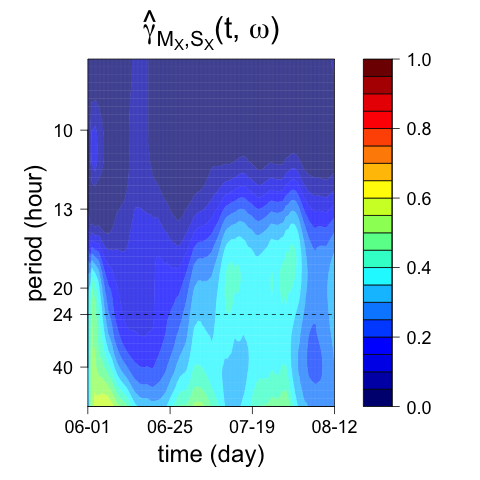}\hfill
    \includegraphics[width = 5cm]{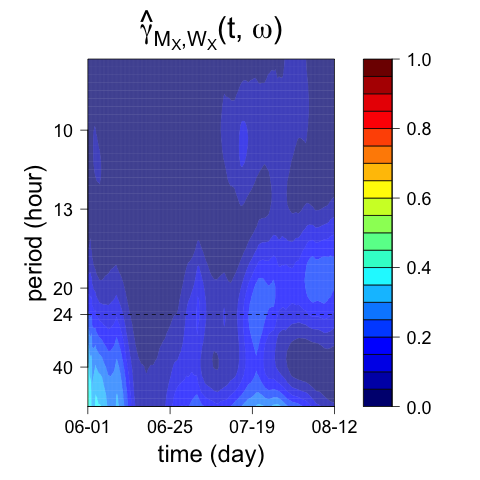}\hfill
    \includegraphics[width = 5cm]{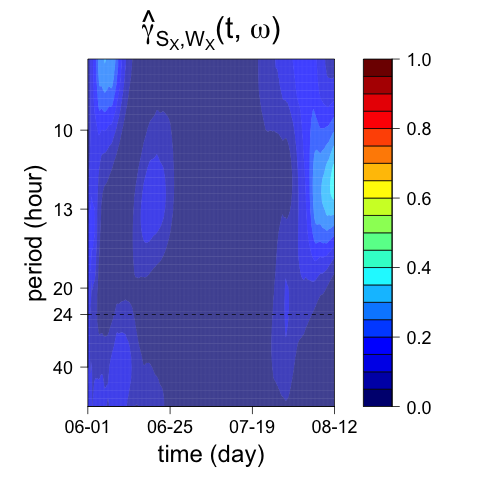}\hfill\\
    \includegraphics[width = 5cm]{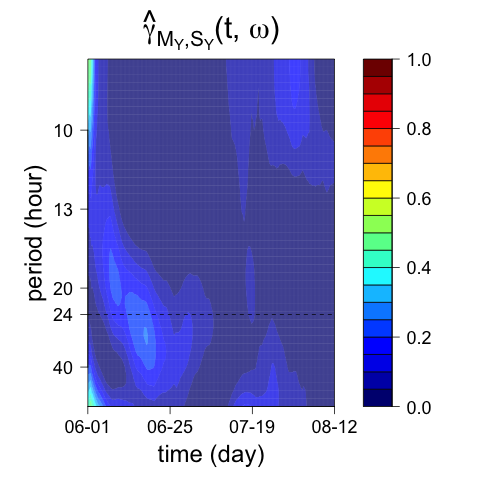}\hfill
    \includegraphics[width = 5cm]{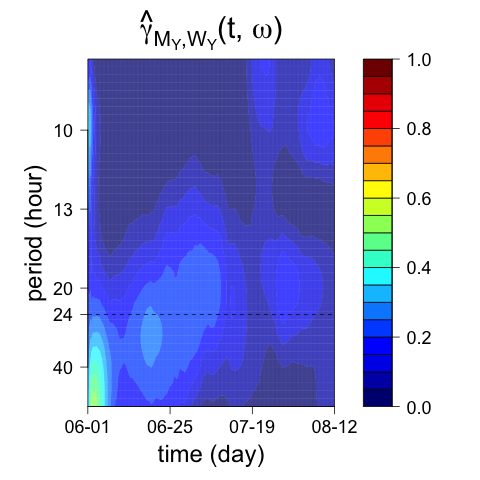}\hfill
    \includegraphics[width = 5cm]{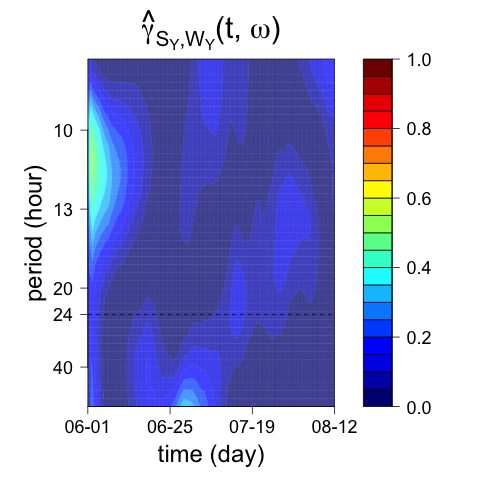}\hfill
    \caption{Top row: The partial coherences between East-West (X) Components and North-South (Y) Components in Monterey, Salinas and Watsonville. Middle row: The partial coherences between Monterey and Salinas, Monterey and Watsonville, as well as Salinas and Watsonville in terms of East-West (X) components. Bottom row: The partial coherences between Monterey and Salinas, Monterey and Watsonville, as well as Salinas and Watsonville in terms of North-South (Y) components.}
    \label{wind_pcoh}
\end{figure}

\end{document}